\documentclass[10pt,journal,epsfig]{IEEEtran}

\usepackage[dvips]{graphicx}
\usepackage{graphicx}
\usepackage{amssymb}
\usepackage{cite}
\usepackage{subfigure}
\usepackage{amsmath}

\usepackage{algorithm}
\usepackage{algorithmic}

\begin{document}


\title{Intelligent Power Control for Spectrum Sharing in Cognitive Radios: A Deep Reinforcement Learning Approach}

\author{Xingjian Li, Jun Fang, Wen Cheng, Huiping Duan, Zhi Chen, and Hongbin Li, ~\IEEEmembership{Senior Member,~IEEE}
\thanks{Xingjian Li, Jun Fang, Wen Cheng and Zhi Chen are with the National Key Laboratory
of Science and Technology on Communications, University of
Electronic Science and Technology of China, Chengdu 611731, China,
Email: JunFang@uestc.edu.cn}
\thanks{Huiping Duan is with the School of Electronic Engineering,
University of Electronic Science and Technology of China, Chengdu
611731, China, Email: huipingduan@uestc.edu.cn}
\thanks{Hongbin Li is
with the Department of Electrical and Computer Engineering,
Stevens Institute of Technology, Hoboken, NJ 07030, USA, E-mail:
Hongbin.Li@stevens.edu}
\thanks{This work was supported in part by the National Science
Foundation of China under Grant 61522104.}}

\maketitle

\IEEEpubid{\begin{minipage}{\textwidth}\ \\[12pt] \centering
~\copyright~2018 IEEE. Personal use of this material is
permitted. Permission
from IEEE must be obtained for all other uses, in any current or future
media, including reprinting/republishing
this material for advertising or
promotional purposes, creating new collective works, for resale or
redistribution to servers or lists, or reuse of any copyrighted
component of this work in other works.
\end{minipage}}

\IEEEpubidadjcol

\begin{abstract}
We consider the problem of spectrum sharing in a cognitive radio
system consisting of a primary user and a secondary user. The
primary user and the secondary user work in a non-cooperative
manner. Specifically, the primary user is assumed to update its
transmit power based on a pre-defined power control policy. The
secondary user does not have any knowledge about the primary
user's transmit power, or its power control strategy. The
objective of this paper is to develop a learning-based power
control method for the secondary user in order to share the common
spectrum with the primary user. To assist the secondary user, a
set of sensor nodes are spatially deployed to collect the received
signal strength information at different locations in the wireless
environment. We develop a deep reinforcement learning-based
method, which the secondary user can use to intelligently adjust
its transmit power such that after a few rounds of interaction
with the primary user, both users can transmit their own data
successfully with required qualities of service. Our experimental
results show that the secondary user can interact with the primary
user efficiently to reach a goal state (defined as a state in
which both users can successfully transmit their data) from any
initial states within a few number of steps.
\end{abstract}


\section{Introduction}
The dramatically increasing demand for spectrum resources requires
new intelligent methods to enhance the spectrum efficiency. Per
the Federal Communications Commission (FCC)
\cite{KolodzyAvoidance02}, the spectrum in general is severely
underutilized with the utilization rate of some bands as low as
$15\%$. In order to improve the spectrum efficiency, the notion of
spectrum sharing with secondary users through cognitive radios is
highly motivated \cite{Haykin05}. Specifically, users from a
secondary network are allowed to access the spectrum owned by
licensed users (also called primary users) without causing harmful
interference.



According to the roles of the primary user, the operation of
spectrum sharing or dynamic spectrum access can be classified into
a passive primary user model and an active primary user model
\cite{WuZhu14}. In many spectrum sharing studies, e.g.
\cite{WangFang10,MitliagkasSidiropoulos11,KimLe08,TadrousSultan11},
it is assumed that the operations of secondary users are
transparent to the primary user so that the primary user does not
need to adapt its transmission parameters. The transparency of
secondary to primary can be accomplished by letting the secondary
user to perform spectrum sensing to explore idle spectrum
\cite{WangFang10} or to strictly control its transmit power such
that the interference to the primary networks is under a desired
threshold \cite{MitliagkasSidiropoulos11,KimLe08,TadrousSultan11}.
However, some works in literature, e.g.
\cite{WuZhu14,SuMatyjas12,ZhuWu13,IslamLiang07}, also considered
an active model in which some (cooperative or non-cooperative)
interaction between the primary user and the secondary user are
allowed to obtain improved transmission performance or economic
compensations. For example, in \cite{WuZhu14}, the spectrum
sharing task is formulated as a Nash bargaining game which
requires interaction between the primary user and the secondary
user to reach a desired equilibrium. Also, in \cite{IslamLiang07},
to achieve spectrum sharing, the primary user and the secondary
user are allowed to interact with each other to update their
respective transmit powers. For the active model, a dynamic power
control strategy is necessary for all users in the network such
that a minimum quality of service (QoS) for successful data
transmission is satisfied for both the primary and the secondary
users.


\IEEEpubidadjcol

Most existing works address this dynamic power control problem
from an optimization perspective. In \cite{GrandhiZander94}, a
distributed constrained power control (DCPC) algorithm was
proposed. Given the signal-to-interference-plus-noise ratio (SINR)
and the required SINR threshold, the DCPC algorithm iteratively
adjusts the transmit power of each transmitter such that all
receivers are provided with their desired QoS requirements. Based
on \cite{GrandhiZander94}, modified approaches with different
constraints or scenarios were developed
\cite{XiaoShroff03,ElBattEphremides04,TadrousSultan10,IslamLiang07,XingMathur07,LeeZeng18}.
Other optimization-based methods were also proposed
\cite{LiuDai15,SenelTekinay17,LiuHong16} in recent years. Besides
optimization-based methods, power allocation from the game
theory's point of view was also studied
\cite{Heikkinen06,ChenZhao13,YangLi15,GaoDuan17}. In
\cite{ChenZhao13}, the power allocation problem was formulated as
a noncooperative game with selfish users, where a sufficient
condition for the existence of a Nash equilibrium was provided,
and a stochastic power adaption with conjecture-based multiagent
Q-learning approach was developed. However, the proposed approach
requires that each user has the knowledge of the channel state
information of every transmitter-receiver pair in the network,
which may be infeasible in practice.



Reinforcement learning \cite{SuttonBarto92}, also known as
Q-learning, has been explored for cognitive radio applications
such as dynamic spectrum access
\cite{BennisNiyato10,Li10,NaparstekCohen17,FuSchaar09,LundenKoivunen11,AlsarhanAgarwal09,WangWen17}.
Using the experience and reward from the environment, users
iteratively optimize their strategy to achieve their goals.
Recently, deep reinforcement learning was introduced and proves
its competence for challenging tasks, say Go and Atari games
\cite{MnihKavukcuoglu13,Mnih15,Silver15}. Unlike conventional
reinforcement learning which is limited to domains with
handcrafted features or low-dimensional observations, agents
trained with deep reinforcement learning are able to learn their
action-value policies directly from high-dimensional raw data such
as images or videos \cite{Silver15}. Also, as to be shown by our
experimental results, deep reinforcement learning can help learn
an effective action-value policy even when the state observations
are corrupted by random noise or measurement errors, while the
conventional Q-learning approach is impractical for such problems
due to the infinite number of states in the presence of random
noise. This characteristic makes deep reinforcement learning
suitable for wireless communication applications whose state
measurements are generally random in nature.





In this paper, we consider a simple cognitive radio scenario
consisting of a primary user and a secondary user. The primary
user and the secondary user work in a non-cooperative manner,
where the primary user adjusts its transmit power based on its own
pre-defined power control policy. The objective is to let the
secondary user learn an intelligent power control policy through
its interaction with the primary user. We assume that the
secondary user does not have any knowledge about the primary
user's transmit power, as well as its power control strategy. To
assist the secondary user, a number of sensors are spatially
deployed to collect the received signal strength (RSS) information
at different locations in the wireless environment. We develop an
intelligent power control policy for the secondary user by
resorting to the deep reinforcement learning approach.
Specifically, the use of deep reinforcement learning, instead of
the conventional reinforcement learning, is to overcome the
difficulty caused by random variations in the RSS measurements.
Our experimental results show that, with the aid of the learned
power control policy, the secondary user can intelligently adjust
its transmit power such that a goal state can be reached from any
initial state within a few number of transition steps.

The rest of the paper is organized as follows. Table
\ref{symbols-table} specifies the frequently-used symbols in this
paper. The system model and the problem formulation are discussed
in Section \ref{sec:system-model}. In Section \ref{sec:DQN}, we
develop a deep reinforcement learning algorithm for power control
for the secondary user. Experimental results are provided in
Section \ref{sec:experiments}, followed by concluding remarks in
Section \ref{sec:conclusions}.

\begin{table}[!t]
\caption{Table of Symbols} \centering
\begin{tabular}{ll}
\noalign{\smallskip} \hline \noalign{\smallskip}
$p_1$ & transmit power of primary user\\
$p_2$ & transmit power of secondary user\\
$h_{ij}$ & channel gain from transmitter $\text{Tx}_i$ to receiver $\text{Rx}_j$\\
$N_i$ & noise power of receiver $\text{Rx}_i$\\
$\text{SINR}_i$ & signal to interference plus noise ratio at receiver $\text{Rx}_i$\\
$\eta_i$ & minimum SINR requirement for receiver $\text{Rx}_i$\\
$N$ & number of sensor nodes\\
$S_n$ & sensor node $n$\\
$P_n^r$ & receive power at sensor node $n$\\
$g_{in}$ & path loss between transmitter $\text{Tx}_i$ and sensor $n$\\
$\sigma_n^2$ & variance of the Gaussian random variable $w_n$\\
$\boldsymbol{s}$ & state of the Markov decision process\\
$a$ & action of the Markov decision process \\
$r$ & reward of the Markov decision process\\
\noalign{\smallskip} \hline
\end{tabular}
\label{symbols-table}
\end{table}

\begin{figure}[!t]
\centering
\includegraphics[width=3.5in]{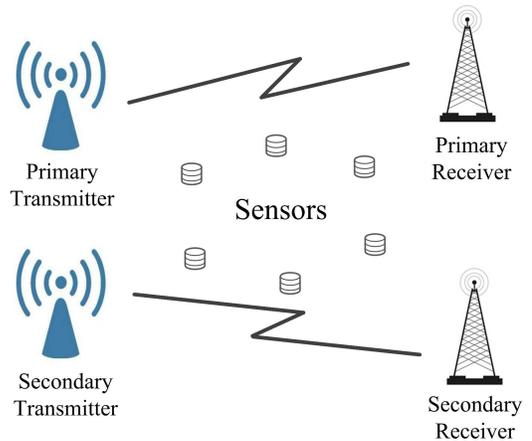}
\caption{A schematic for spectrum sharing in cognitive radio
networks.} \label{system-model}
\end{figure}

\section{System Model} \label{sec:system-model}
Consider a cognitive radio network consisting of a primary user
and a secondary user, where the secondary user aims to share a
common spectrum resource with the primary user, without causing
harmful interference to the primary user. The primary user
consists of a primary transmitter ($\text{Tx}_1$) and a primary
receiver ($\text{Rx}_1$), and the secondary user consists of a
secondary transmitter ($\text{Tx}_2$) and a secondary receiver
($\text{Rx}_2$), see Fig. \ref{system-model}. In our setup, we
assume that the primary user and the secondary user are working in
a non-cooperative way, in which the primary user is unaware of the
existence of the secondary user, and adjusts its transmit power
based on its own power control policy. Nevertheless, it should be
noted that since the power control policy for the primary user is
dependent on the environment (cf. (\ref{power-control-policy-1})
and (\ref{power-control-policy-2})), the action taken by the
secondary user at the current time will affect the primary user's
next move in an implicit way. There is also no communication
between the primary network and the secondary network. Thus the
secondary user has no knowledge about the primary user's transmit
power and its power control policy. For simplicity, we, at this
point, assume that the primary user and the secondary user
synchronously update their respective transmit power and the
transmit power is adjusted on a time framed basis. We will show
later our proposed scheme also works when the synchronous
assumption does not hold.



The objective here is to help the secondary user learn an
efficient power control policy such that, after a few rounds of
power adjustment, both the primary user and the secondary user are
able to transmit their data successfully with required QoSs.
Clearly, this task cannot be accomplished if the secondary user
only knows its own transmit power. To assist the secondary user, a
set of sensor nodes are employed to measure the received signal
strength (RSS) at different locations in the wireless environment.
The RSS measurements are related to both users' transmit power,
thus revealing the state information of the system. We assume that
the RSS information is accessible to the secondary user. Note that
collecting the RSS information from spatially distributed sensor
nodes is a basic requirement for many applications, e.g. source
localization \cite{TomicBeko15}. For our problem, each node only
needs to report the RSS information once per time frame, which
involves a low data rate. Therefore some conventional technologies
such as the Zigbee \cite{YickMukherjee08} which delivers
low-latency communication for wireless mesh networks can be
employed to provide timely feedback of the RSS information from
sensor nodes to the secondary user.


For both the primary user and the secondary user, the QoS is
measured in terms of the SINR. Let $p_1$ and $p_2$ denote the
transmit power of the primary user and the secondary user,
respectively. The SINR for the $i$th receiver is given as
\begin{align}
\text{SINR}_i = \frac{|h_{ii}|^2p_i}{\sum_{j\neq
i}|h_{ji}|^2p_j+N_i} \quad i=1,2
\end{align}
where $h_{ij}$ denotes the channel gain from the transmitter
$\text{Tx}_i$ to the receiver $\text{Rx}_j$, and $N_i$ is the
noise power at the receiver $\text{Rx}_i$. We assume that the
primary receiver and the secondary receiver have to satisfy a
minimum SINR requirement for successful reception, i.e.
$\text{SINR}_i\geq \eta_i, i=1,2$.

To meet the QoS requirement, the primary user is supposed to
adaptively adjust its transmit power based on its own power
control policy. In this paper, two different power control
strategies are considered for the primary user. Note that our
proposed method also works if the primary user adopts other power
control policies. For the first strategy, the transmit power of
the primary user is updated according to the classical power
control algorithm \cite{GrandhiZander94}
\begin{align}
p_1(k+1) = D\left(\frac{\eta_1p_1(k)}{\text{SINR}_1(k)}\right)
\label{power-control-policy-1}
\end{align}
where $\text{SINR}_1(k)$ denotes the SINR measured at the primary
receiver at the $k$th time frame, $p_1(k)$ denotes the transmit
power at the $k$th time frame, here we assume that the transmit
power is adjusted on a time framed basis. $D(\cdot)$ is a
discretization operation which maps continuous-valued levels into
a set of discrete values
\begin{align}
\mathcal{P}_1\triangleq\{p^p_1,\ldots,p^p_{L_1}\}
\end{align}
where $p^p_1\leq\ldots\leq p^p_{L_1}$. More precisely, we let
$D(x)$ equal the nearest discrete level that is no less than $x$
and let $D(x)=p_{L_1}^p$ if $x>p_{L_1}^p$. For the second power
control strategy, suppose the transmit power at the $k$th time
frame is $p_1(k)=p_j^p$, where $p_j^p\in \mathcal{P}_1$. The
transmit power of the primary user is updated according to
\begin{align}
p_1(k+1) =
\begin{cases}
p_{j+1}^p   &  \text{if} \ p_j^p \leq \tau\leq p_{j+1}^p \ \text{and} \ j+1\leq L_1 \\
p_{j-1}^p  &  \text{if} \ \tau\leq p_{j-1}^p \ \text{and} \ j-1\geq 1 \\
p_j^p  &  \text{otherwise}
\end{cases} \label{power-control-policy-2}
\end{align}
where $\tau\triangleq \eta_1 p_1(k)/\text{SINR}_{1}(k)$. We see
that compared to (\ref{power-control-policy-1}), the power control
policy (\ref{power-control-policy-2}) has a more conservative
behavior: it updates its transmit power in a stepwise manner.
Specifically, it increases its power (by one step) when
$\text{SINR}_1(k)\leq \eta_1$ and $\hat{\eta}\geq\eta_1$, and
decreases its power (by one step) when $\text{SINR}_1(k)\geq
\eta_1$ and $\hat{\eta}\geq\eta_1$; otherwise it will stay on the
current power level. Here
$\hat{\eta}\triangleq\text{SINR}_1(k)p_1(k+1)/p_1(k)$ is the
`predicted' SINR at the $(k+1)$th time frame.

Suppose $N$ sensors are deployed to spatially sample the RSS
information. Let $S_n$ denote node $n$, and $P_{n}^r(k)$ denote
the receive power at sensor $n$ at the $k$th frame. In our paper,
the following model is used to simulate the state (i.e. RSS)
observations
\begin{align}
P_{n}^r(k)=p_1(k)g_{1n}+p_2(k)g_{2n}+w_n(k)
\label{state-observation-model}
\end{align}
where $p_1(k)$ and $p_2(k)$ represent the transmit power of the
primary user and the secondary user, respectively, $g_{1n}$
denotes the path loss between the primary transmitter and sensor
$n$, $g_{2n}$ denotes the path loss between the secondary
transmitter and sensor $n$, and $w_{n}(k)$, a zero mean Gaussian
random variable with variance $\sigma_n^2$, is used to account for
the random variation caused by shadowing effect and estimation
errors. For free-space propagation, according to the Friis law
\cite{Rappaport02}, $g_{1n}$ and $g_{2n}$ are respectively given
by
\begin{align}
g_{1n}=\left(\frac{\lambda}{4\pi d_{1n}}\right)^2 \quad
g_{2n}=\left(\frac{\lambda}{4\pi d_{2n}}\right)^2
\end{align}
where $\lambda$ is the signal wavelength, $d_{1n}$ ($d_{2n}$)
denotes the distance between the primary (secondary) transmitter
and node $n$.

We also assume that the transmit power of the secondary user is
chosen from a finite set
\begin{align}
\mathcal{P}_2\triangleq\{p^s_1,\ldots,p^s_{L_2}\}
\end{align}
where $p^s_1\leq\ldots\leq p^s_{L_2}$. The objective of the
secondary user is to learn how to adjust its transmit power based
on the collected RSS information $\{P_{n}^r(k)\}_n$ at each time
frame such that after a few rounds of power adjustment, both the
primary user and the secondary user can meet their respective QoS
requirements for successful data transmissions. Note that we
suppose there exists at least a pair of transmit power
$\{p^p_{l_1},p^s_{l_2}\}$ such that the primary receiver and the
secondary receiver satisfy their respective QoS (SINR)
requirements, i.e. $\text{SINR}_i\geq\eta_i,i=1,2$.

\begin{figure}[!t]
\centering
\includegraphics[width=3.5in]{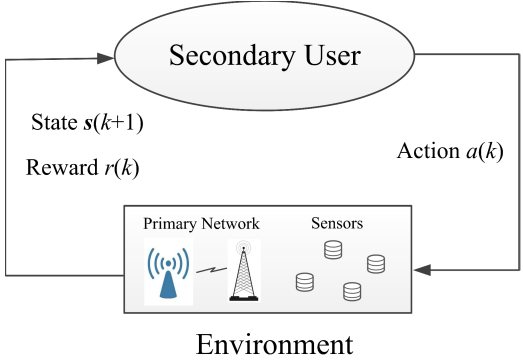}
\caption{Interaction between the secondary user and the
environment (i.e. the primary user).} \label{interaction}
\end{figure}

\section{A Deep Reinforcement Learning Approach for Power Control}
\label{sec:DQN} We see that the secondary user, at each time
frame, has to take an action (i.e. choose a transmit power from a
pre-specified power set $\mathcal{P}_2$) based on its current
state
\begin{align}
\boldsymbol{s}(k)\triangleq
\left[P_{1}^r(k)\phantom{0}\ldots\phantom{0}P_{N}^r(k)\right]^T
\end{align}
This power control process is essentially a Markov decision
process (MDP) because after the decision maker (i.e. the secondary
user) chooses any action $a(k)=p_2(k+1)$ in state
$\boldsymbol{s}(k)$, the process will move into a new state
$\boldsymbol{s}(k+1)$ which depends on the current state
$\boldsymbol{s}(k)$ and the decision maker's action $a(k)$, and
given $\boldsymbol{s}(k)$ and $a(k)$, the next state is
conditionally independent of all previous states and actions.
Also, after moving into a new state, the decision maker will
receive a corresponding reward $r(k)\triangleq
r(\boldsymbol{s}(k),a(k))$ which can be defined as
\begin{align}
r(k)\triangleq
\begin{cases}
10   &  \text{if} \ \text{SINR}_1(k+1)\geq\eta_1 \ \text{and} \ \text{SINR}_2(k+1)\geq\eta_2\\
0  &  \text{otherwise} \nonumber
\end{cases}
\end{align}
The interaction between the secondary user and the environment is
shown in Fig. \ref{interaction}. Note that here the decision maker
(secondary user) is assumed to know whether the transmission
between the primary transmitter and the primary receiver is
successful or not. In practice, such knowledge may be obtained by
monitoring an acknowledgment signal sent by the primary receiver
to indicate successful receipt of a transmission from the primary
transmitter.



The core problem of MDPs is to learn a ``policy'' for the decision
maker: a function $\pi$ that specifies the action
$\pi(\boldsymbol{s})$ that the decision maker will choose when in
state $\boldsymbol{s}$. More precisely, the goal of the secondary
user is to learn a policy $\pi$ for selecting its action $a(k)$
based on the current state $\boldsymbol{s}(k)$ in a way that
maximizes a discounted cumulative reward which is defined as
\cite{SuttonBarto92}
\begin{align}
  V^{\pi}\left(\boldsymbol{s}(k)\right) &\triangleq \sum_{i=k}^{T'} \gamma^{i-k} r(i)
\end{align}
where $\gamma$ is the discount factor and $T'$ denotes the time
frame at which the goal state is reached. For our problem, the
goal state is defined as a state in which $\text{SINR}_i(k) \geq
\eta_i, i=1,2$. Thus, the task becomes learning an optimal policy
$\pi^*$ that maximizes $V^{\pi}$, i.e.
\begin{align}
\pi^* &= \mathop{\arg\max}_{\pi}{V^{\pi}(\boldsymbol{s})} \quad
\forall \boldsymbol{s}
  \label{max_pi}
\end{align}
Directly learning $\pi^*$ is difficult. In reinforcement learning,
Q-learning provides an alternative approach to solve
(\ref{max_pi}) \cite{WatkinsDayan92}. Instead of learning $\pi^*$,
an action-value (also known as Q) function is introduced to
evaluate the expected discounted cumulative reward after taking
some action $\boldsymbol{a}$ in a given state $\boldsymbol{s}$.
When such an action-value function is learned, the optimal policy
can be constructed by simply selecting the action with the highest
value in each state. The basic idea behind the Q-learning and many
other reinforcement learning algorithms is to iteratively update
the action-value function according to a simple value iteration
update rule
\begin{align}
Q(\boldsymbol{s},a)= r(\boldsymbol{s},a) + \gamma
\max_{a'}{Q(\boldsymbol{s}',a')} \label{Bellman-equation}
\end{align}
The above update rule is also known as the Bellman equation
\cite{Bellman03}, in which $\boldsymbol{s}'$ is the state
resulting from applying action $a$ to the current state
$\boldsymbol{s}$. It has been proved that the value iteration
algorithm (\ref{Bellman-equation}) converges to the optimal
action-value function, which is defined as the maximum expected
discounted cumulative reward by following any policy, after taking
some action $\boldsymbol{a}$ in a given state $\boldsymbol{s}$.
For the Q-learning, the number of states is finite and the
action-value function is estimated separately for each state, thus
leading to a Q-table or a Q-matrix, with its rows representing the
states and its columns representing the possible actions. After
the Q-table converges, one can select an action $a$ which has the
largest value of $Q(\boldsymbol{s},a)$ as the optimal action in
state $\boldsymbol{s}$.

Unfortunately, due to the random variation in the RSS measurement,
the value of $\boldsymbol{s}$ is continuous. As a result, the
Q-learning approach is impractical for our problem since we could
have an infinite number of states. To overcome this issue, we
resort to the deep Q-network (DQN) proposed in \cite{Mnih15}.
Unlike the conventional Q-learning method that generates a finite
action-value table, for the DQN, the table is replaced by a deep
neural network $Q(\boldsymbol{s},a;\boldsymbol{\theta})$ to
approximate the action-value function, where $\boldsymbol{\theta}$
denotes the weights of the Q-network. Specifically, given an input
$\boldsymbol{s}$, the deep neural network yields an
$L_2$-dimensional vector, with its $i$th entry representing the
estimated value for choosing the action $a=p^{s}_i$ from
$\mathcal{P}_2$.

The training data used to train the Q-network are generated as
follows. Given $\boldsymbol{s}(k)$, at iteration $k$, we either
explore a randomly selected action with probability
$\varepsilon_k$, or select an action $a(k)$ which has the largest
output $Q(\boldsymbol{s}(k),a(k);\boldsymbol{\theta}_0)$, where
$\boldsymbol{\theta}_0$ denotes the parameters for the current
iteration. After taking the action $a(k)$, the secondary user
receives a reward $r(k)$ and observes a new state
$\boldsymbol{s}(k+1)$. This transition
$d(k)\triangleq\{\boldsymbol{s}(k),a(k),r(k),\boldsymbol{s}(k+1)\}$
is stored in the replay memory $D$. The training of the Q-network
begins when $D$ has collected a sufficient number of transitions,
say $O=300$ transitions. Specifically, we randomly select a
minibatch of transitions $\{d(i)|i\in\Omega_k\}$ from $D$, and the
Q-network can be trained by adjusting the parameters
$\boldsymbol{\theta}$ such that the following loss function is
minimized
\begin{align}
  L(\boldsymbol{\theta}) \triangleq \frac{1}{|\Omega_k|}
  \sum_{i\in\Omega_k} \left(Q'(i)-Q(\boldsymbol{s}(i),a(i);\boldsymbol{\theta})\right)^2
  \label{loss-function}
\end{align}
in which $\Omega_k$ is the index set of the random minibatch used
at the $k$th iteration, and $Q'(i)$ is a value estimated via the
Bellman equation by using parameters from the current iteration,
i.e.
\begin{align}
Q'(i) = r(i)+\gamma \max_{a'}
Q(\boldsymbol{s}(i+1),a';\boldsymbol{\theta}_0) \quad  \forall
i\in \Omega_k \label{Q'}
\end{align}
Note that unlike traditional supervised learning, the targets for
DQN learning is updated as the weights $\boldsymbol{\theta}$ are
refined. For clarity, we summarize our proposed DQN training
algorithm in Algorithm \ref{DQN-training}.




\begin{figure*}[!t]
\centering
\includegraphics[width=7in]{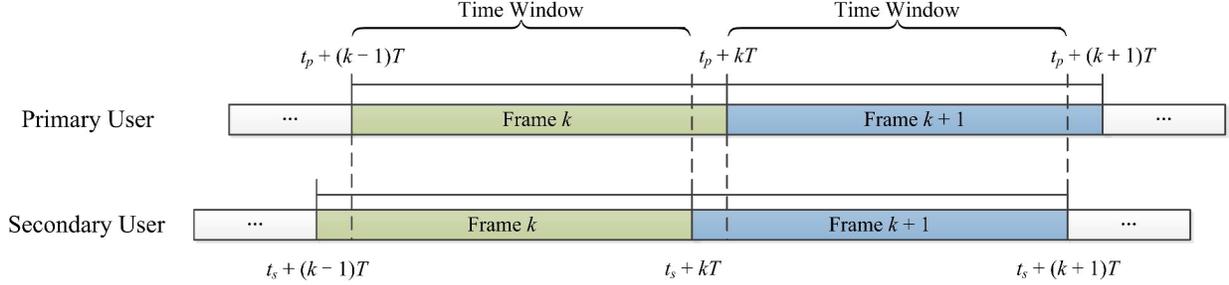}
\caption{Asynchronous update of the transmit power for the primary
user and the secondary user.} \label{asynchronous-system}
\end{figure*}

\begin{algorithm}[!t]
\caption{DQN Training for Power Control}
\begin{algorithmic}
\STATE {
Initialize replay memory $D$ with buffer capacity $O$\\
Initialize network $Q(\boldsymbol{s},a,\boldsymbol{\theta})$
with random weights $\boldsymbol{\theta}=\boldsymbol{\theta}_0$\\
Initialize $p_1(1)$ and $p_2(1)$, then obtain $\boldsymbol{s}(1)
$\FOR{$k=1,K$}
\STATE Update $p_1(k+1)$ via the primary user's power control
strategy (\ref{power-control-policy-1}) or
(\ref{power-control-policy-2}) \STATE With probability
$\varepsilon_k$ select a random action $a(k)$, otherwise select
$a(k)=\max_a Q(\boldsymbol{s}(k),a;\boldsymbol{\theta}_0)$ \STATE
Obtain $\boldsymbol{s}(k+1)$ via the random observation model
(\ref{state-observation-model}) and observe reward $r(k)$ \STATE
Store transition
$d(k)\triangleq\{\boldsymbol{s}(k),a(k),r(k),\boldsymbol{s}(k+1)\}$
in $D$ \IF {$k\geq O$} \STATE Sample a random minibatch of
transitions $\{d(i)|i\in\Omega_k\}$ from $D$, where the indexes in
$\Omega_k$ are uniformly chosen at random \STATE Update
$\boldsymbol{\theta}$ by minimizing the loss function
(\ref{loss-function}), where targets $Q'(i)$ are given by
(\ref{Q'}) \STATE Set
$\boldsymbol{\theta}_0=\mathop{\arg\min}_{\boldsymbol{\theta}}L(\boldsymbol{\theta})$
\ENDIF \IF {$\boldsymbol{s}(k)$ is a goal state} \STATE Initialize
$p_1(k+1)$ and $p_2(k+1)$, then obtain $\boldsymbol{s}(k+1)$
\ENDIF
\ENDFOR }
\end{algorithmic}
\label{DQN-training}
\end{algorithm}

After training, the secondary user can choose the action which
yields the largest estimated value
$Q(\boldsymbol{s},a,\boldsymbol{\theta}^*)$. For clarity, the
proposed DQN-based power control scheme for the secondary user is
summarized in Algorithm \ref{DQN-based-Power-Control}. We would
like to point out that during the DQN training process, the
secondary user requires the knowledge of whether the QoS
requirements for the primary user and the secondary user are
satisfied. Nevertheless, after the DQN is trained, the secondary
user only needs the feedback from sensors to decide its next
transmit power.

We discuss the convergence issue of the proposed power control
policy. Suppose $\boldsymbol{s}$ is a goal state. If the transmit
power of the secondary user remains unchanged, then it is easy to
show that the next state $\boldsymbol{s}'$ is also a goal state,
whichever of (\ref{power-control-policy-1}) and
(\ref{power-control-policy-2}) is chosen for the primary user to
update its transmit power. On the other hand, the secondary user
will eventually learn to choose a transmit power such that the
next state $\boldsymbol{s}'$ remains a goal state. Therefore we
can conclude that once $\boldsymbol{s}$ reaches a goal state, it
will stay at the goal state until the data transmission is over.
Suppose the goal state is lost due to the discontinuity of data
transmission, and the secondary user wants to restart a new
transmission. In this case, learning is no longer required. The
secondary user can select its transmit power according to the
learned power control policy.

In our previous discussion, we assume that the primary user and
the secondary user synchronously update their respective transmit
power. Nevertheless, we would like to point out that the
synchronous assumption is not necessarily required by our proposed
scheme. Suppose the time frames between the primary user and the
secondary user are not strictly synchronized (see Fig.
\ref{asynchronous-system}). Both the primary user and the
secondary user update their transmit power at the beginning of
their respective time frames, that is, the primary user adjusts
its transmit power at time $t_p,t_p+T,t_p+2T,\ldots$, and the
secondary user updates its transmit power at time $t_s,t_s+T,
t_s+2T,\ldots$, where $T$ denotes the duration of each frame.
Without loss of generality, we assume $T>t_p-t_s>0$. Clearly, our
intelligent power control scheme would function the same as in the
synchronous case if both the primary user and the secondary user
perform their respective tasks, i.e. gather necessary information
(i.e. $\text{SINR}_1(k)$ for the primary user, $\text{SINR}_1(k)$,
$\text{SINR}_2(k)$, and $\boldsymbol{s}(k)$ for the secondary
user) and make decisions during the time window
$[t_p+(k-1)T,t_s+kT]$.



\begin{algorithm}[!t]
\caption{DQN-based Power Control Strategy}
\begin{algorithmic}
\STATE { Initialize $p_2(1)$, then obtain $\boldsymbol{s}(1)$
\FOR{$k =1,K$} \STATE Select $a(k)=\max_a
Q(\boldsymbol{s}(k),a;\boldsymbol{\theta}^*)$ \STATE Obtain
$\boldsymbol{s}(k+1)$ \ENDFOR }
\end{algorithmic}
\label{DQN-based-Power-Control}
\end{algorithm}



\begin{figure*}[!t]
\centering \subfigure[Loss function vs. the number of
iterations.]{\includegraphics[width=2.3in]{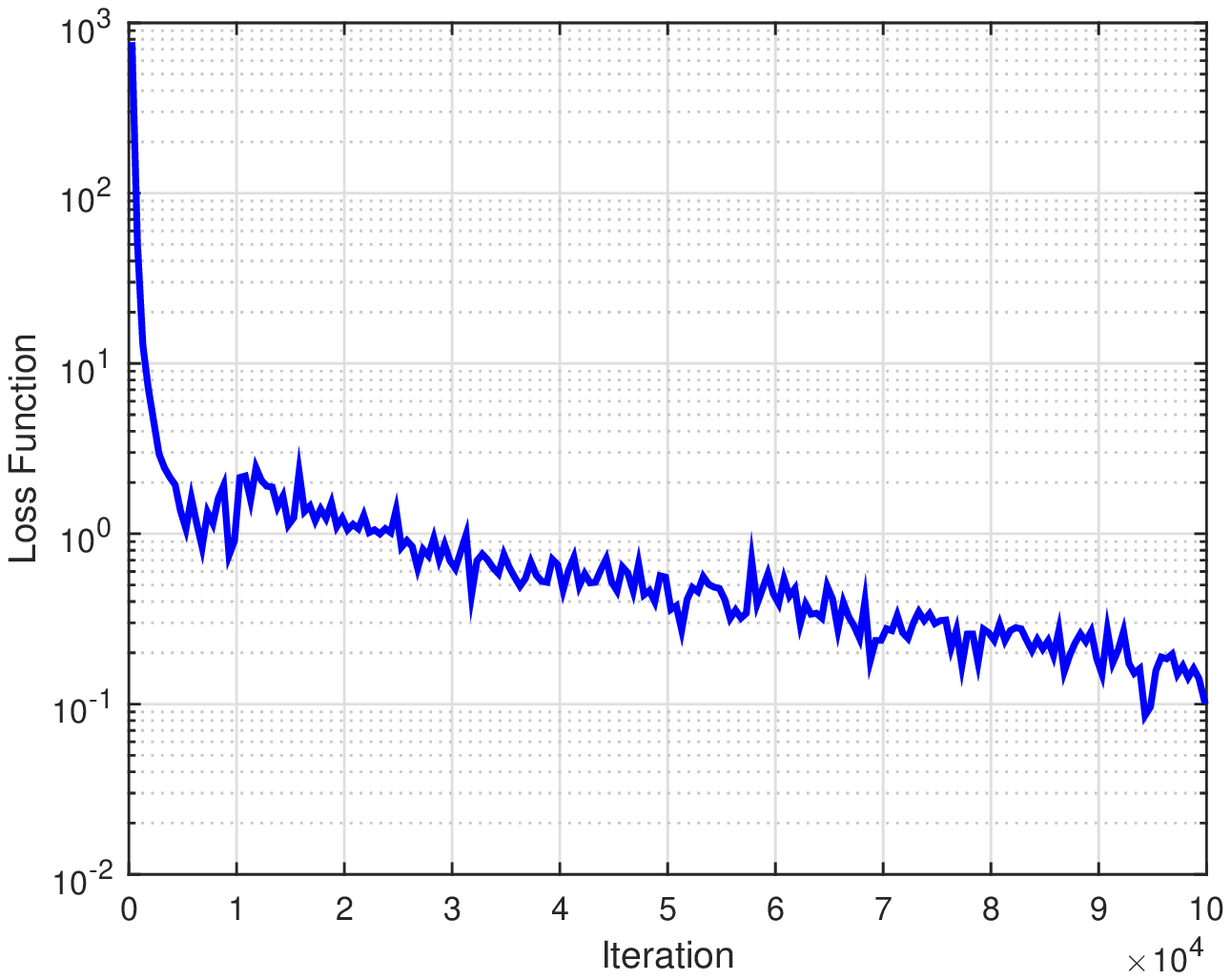}} \hfil
\subfigure[Success rate vs. the number of
iterations.]{\includegraphics[width=2.3in]{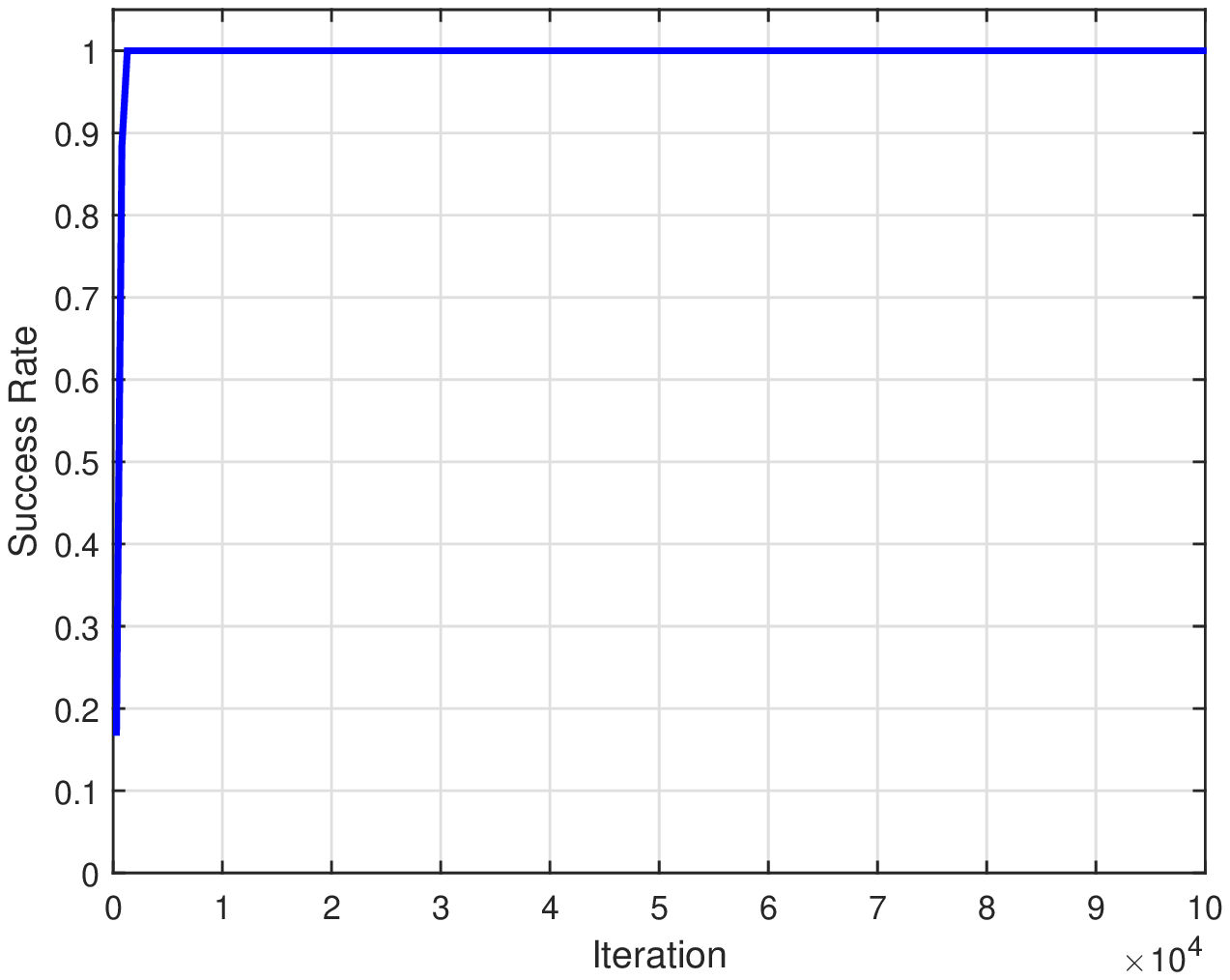}} \hfil
\subfigure[Average number of transition steps vs. the number of
iterations.]{\includegraphics[width=2.3in]{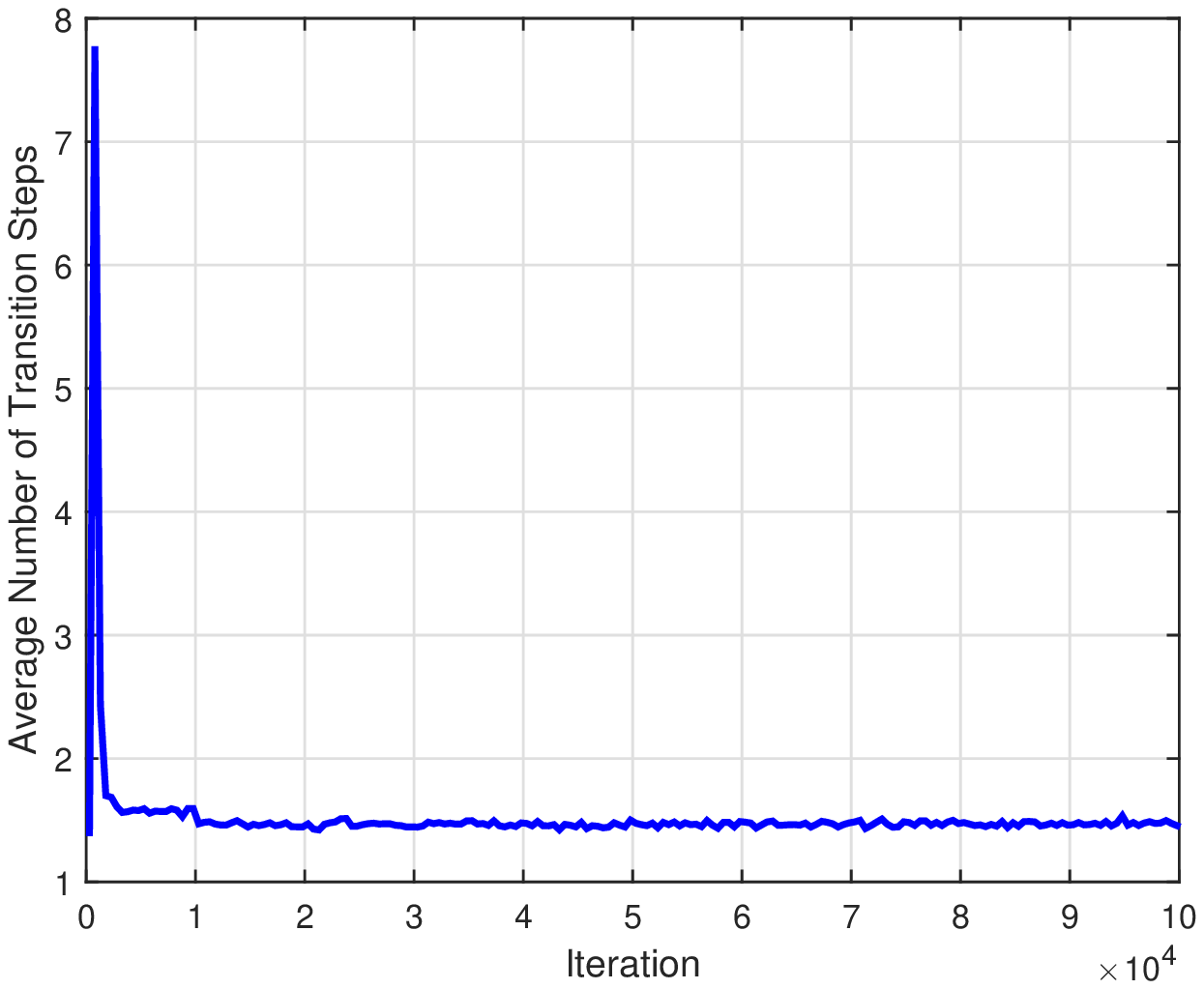}}
\caption{Loss function, success rate, and average number of
transition steps vs. the number of iterations $k$ used for
training, where $N=10$, $\sigma_n = (p_1^pg_{1n}+
p_1^sg_{2n})/10$.} \label{fig11}
\end{figure*}

\begin{figure*}[!t]
\centering \subfigure[Loss function vs. the number of
iterations.]{\includegraphics[width=2.3in]{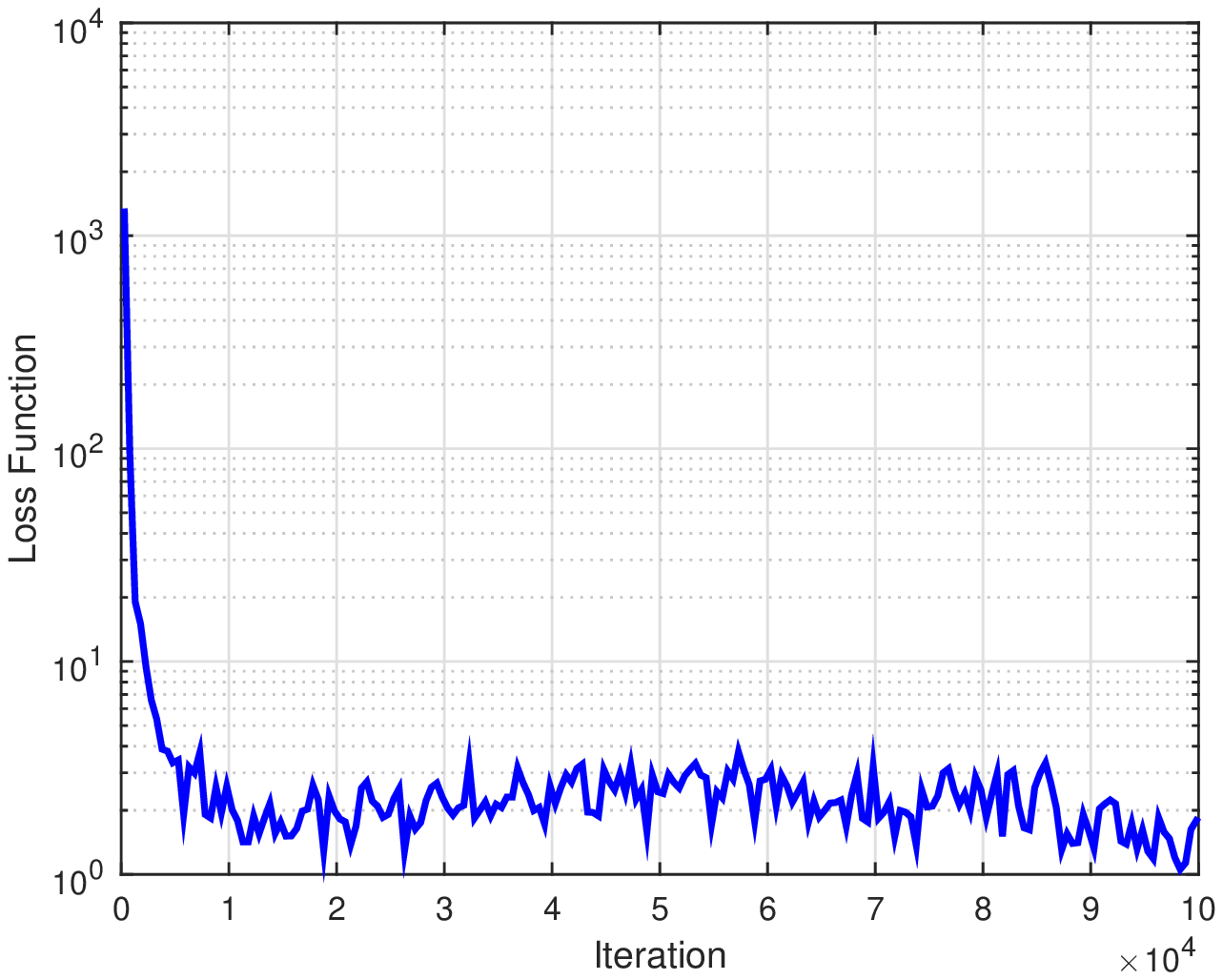}}\hfil
\subfigure[Success rate vs. the number of
iterations.]{\includegraphics[width=2.3in]{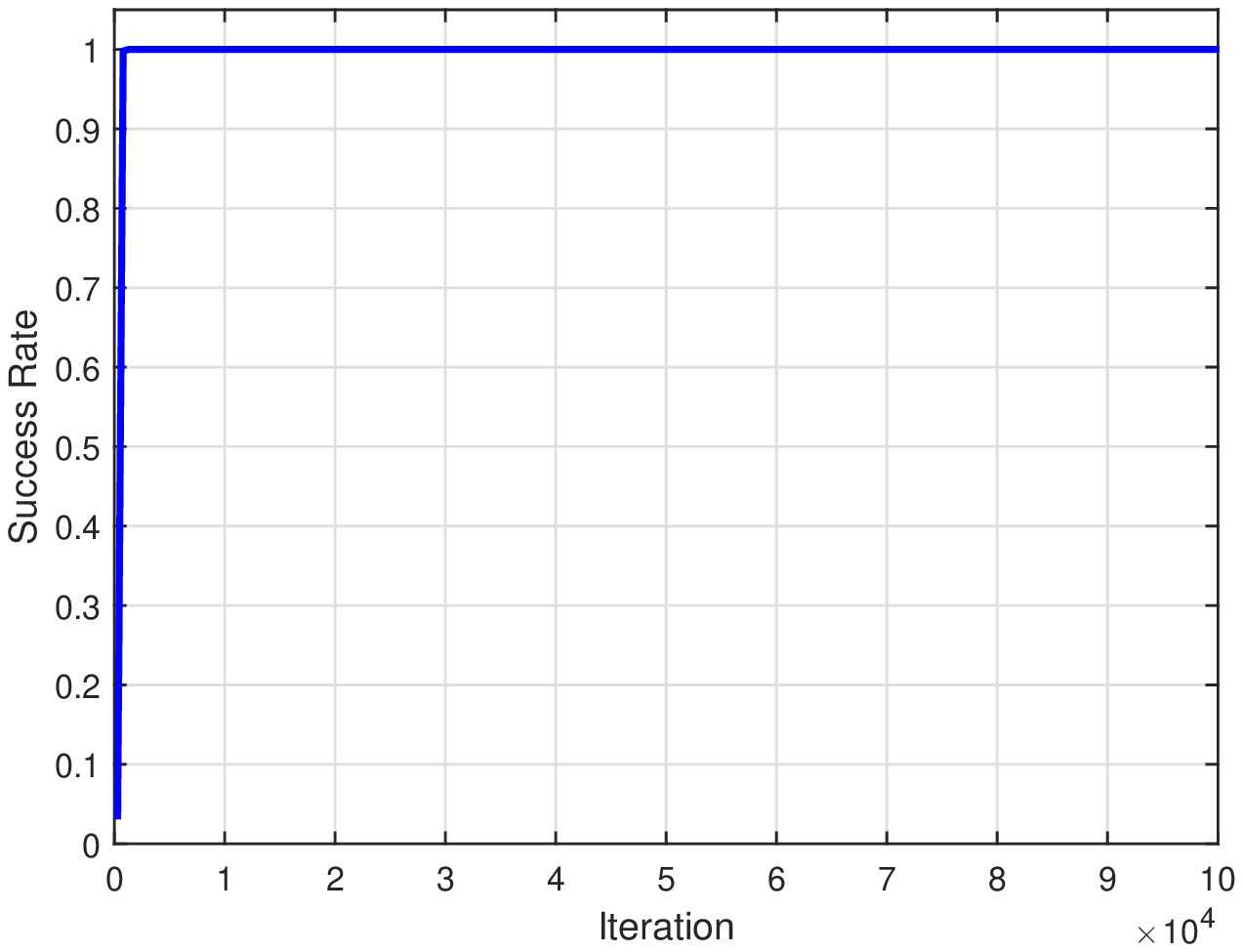}} \hfil
\subfigure[Average number of transition steps vs. the number of
iterations.]{\includegraphics[width=2.3in]{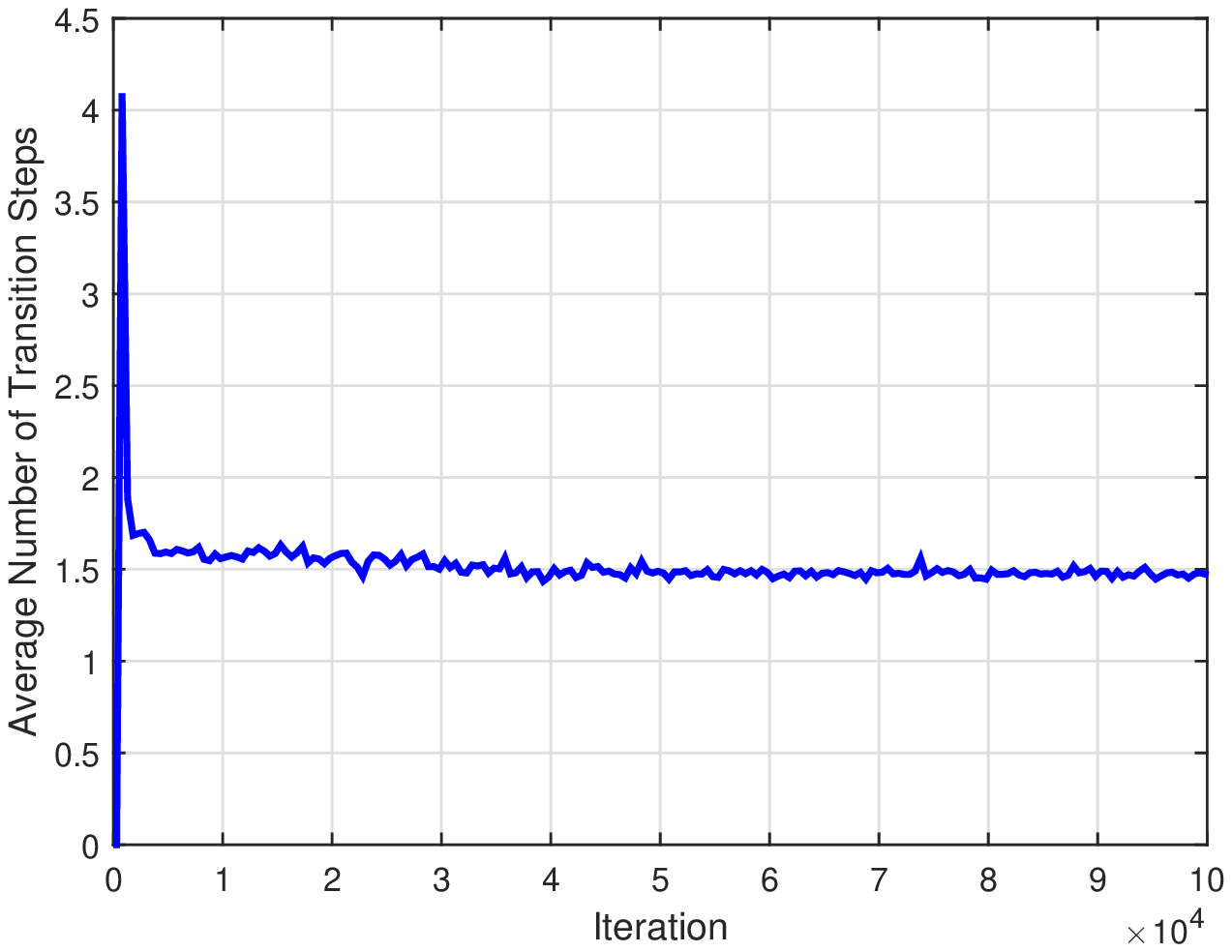}}
 \caption{Loss
function, success rate, and average number of transition steps vs.
the number of iterations $k$ used for training, where $N=10$,
$\sigma_n = (p_1^pg_{1n}+ p_1^sg_{2n})/3$. } \label{fig12}
\end{figure*}

\begin{figure*}[!t]
\centering \subfigure[Loss function vs. the number of
iterations.]{\includegraphics[width=2.3in]{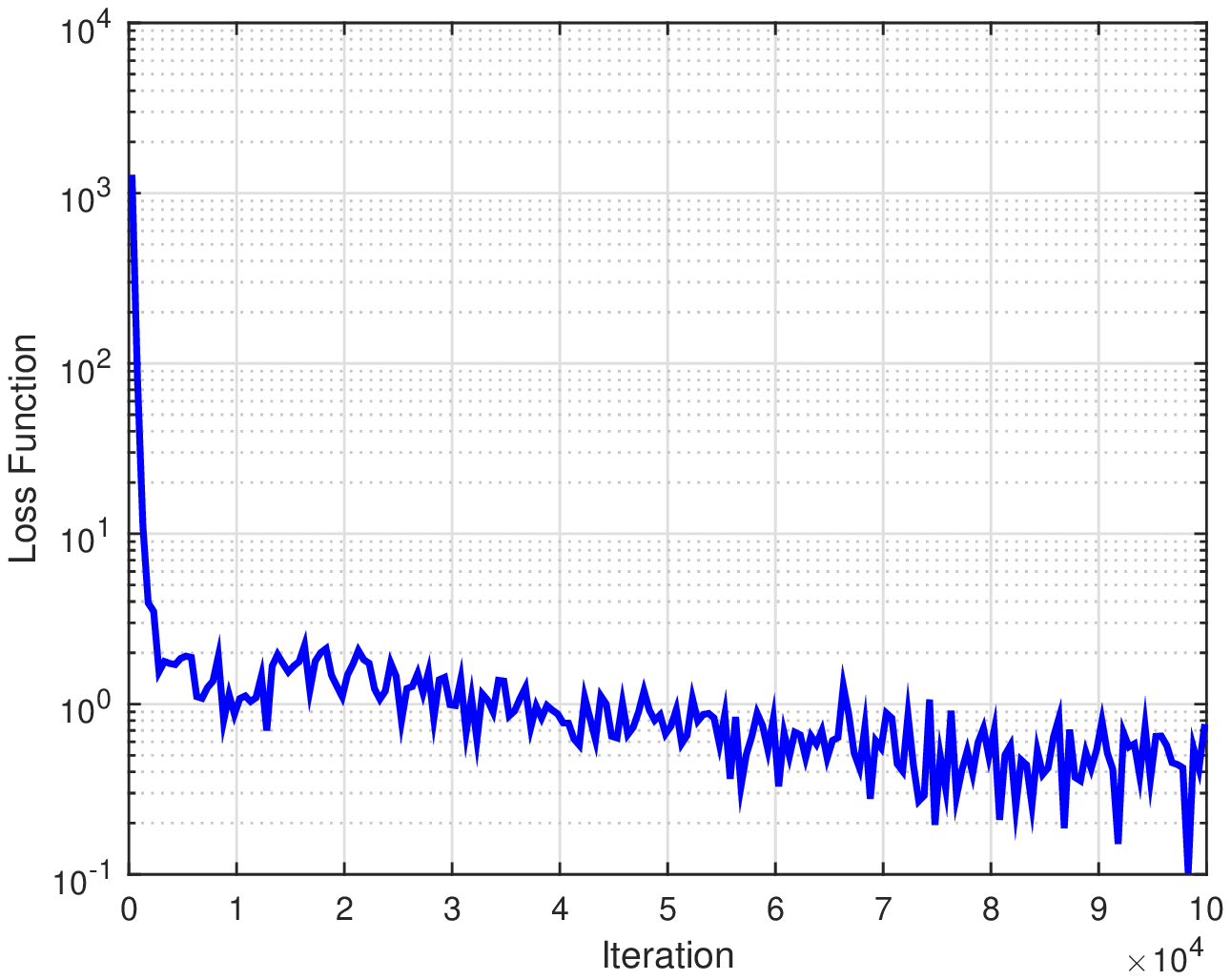}}\hfil
\subfigure[Success rate vs. the number of
iterations.]{\includegraphics[width=2.3in]{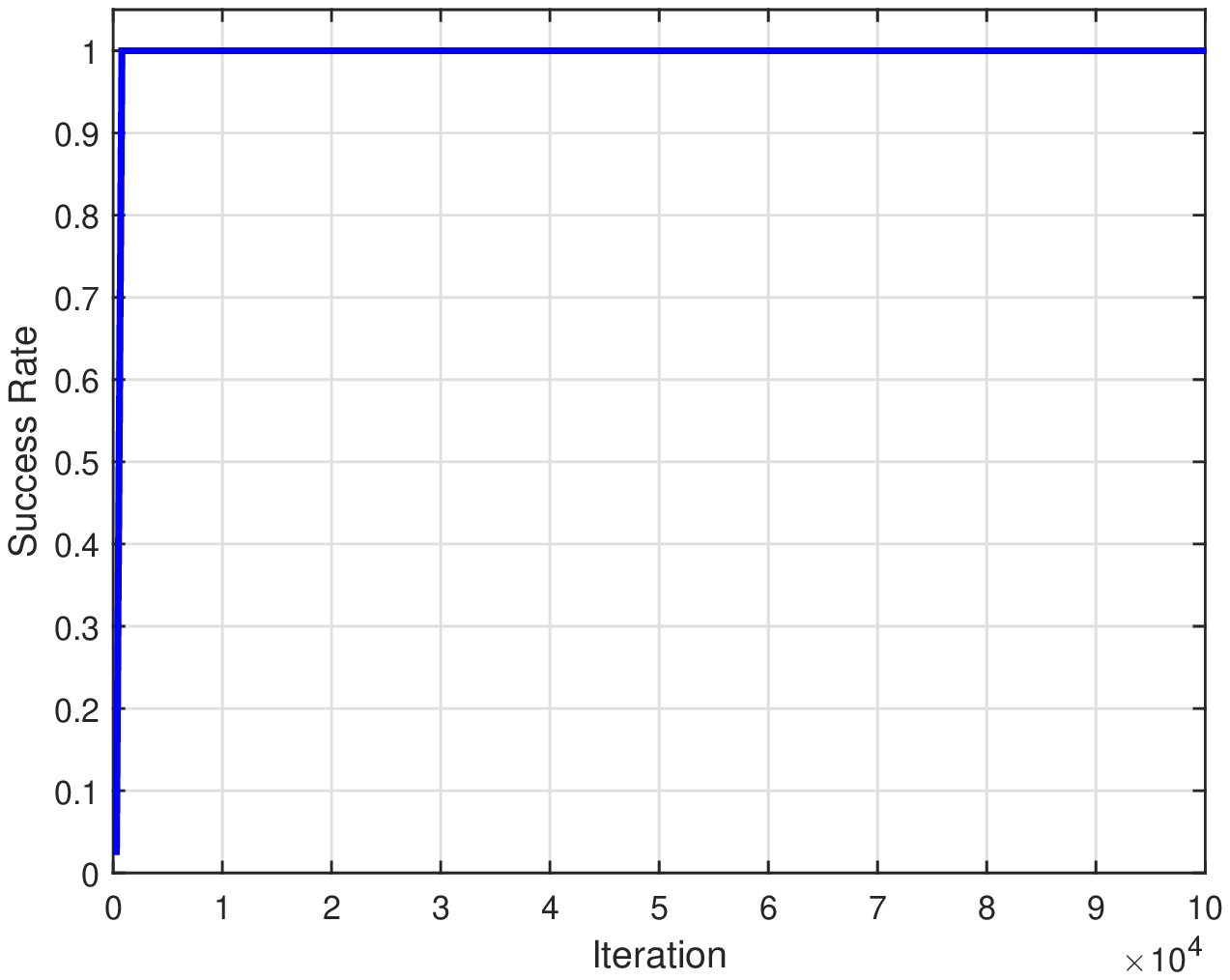}} \hfil
\subfigure[Average number of transition steps vs. the number of
iterations.]{\includegraphics[width=2.3in]{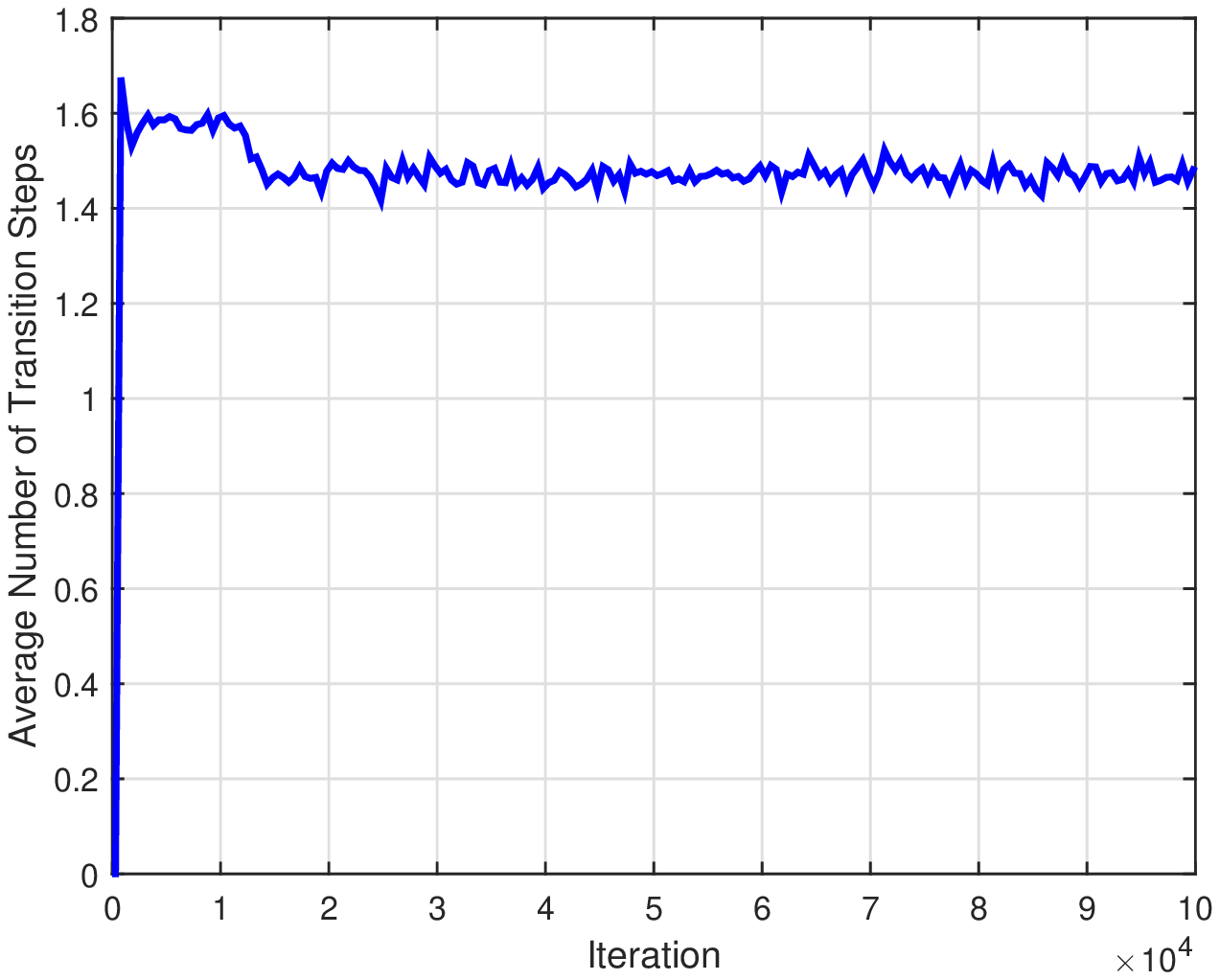}}
 \caption{Loss
function, success rate, and average number of transition steps vs.
the number of iterations $k$ used for training, where $N=3$,
$\sigma_n = (p_1^pg_{1n}+ p_1^sg_{2n})/10$.} \label{fig13}
\end{figure*}

\section{Experimental Results} \label{sec:experiments}
We now carry out experiments to illustrate the performance of our
proposed DQN-based power control algorithm\footnote{Codes are
available at
http://www.junfang-uestc.net/codes/DQN-power-control.rar}. In our
experiments, the transmit power (in Watt) of both the primary user
and the secondary user is chosen from a pre-defined set
$\mathcal{P}_1=\mathcal{P}_2=\{0.05,0.1,\ldots,0.4\}$, and the
noise power at $\text{Rx}_1$ and $\text{Rx}_2$ is set to
$N_1=N_2=0.01$W. For simplicity, the channel gains from the
primary/secondary transmitter to the primary/secondary receivers
are assumed to be $h_{ij}=1,\forall i,j$. The minimum SINR
requirements for successful reception for the primary user and the
secondary user are set to $\eta_1=1.2$, $\eta_2=0.7$,
respectively. It can be easily checked that there exists a pair of
transmit power $\{p_1,p_2\}$ which ensures that the QoSs of the
primary user and the secondary user are satisfied. Also, a total
number of $N$ sensors are employed to collect the RSS information
to assist the secondary user to learn a power control policy. The
distance $d_{ij}$ between the transmitter $\text{Tx}_i$ and the
sensor node $S_j$ is uniformly distributed in the interval $[100,
300]$ (in meters).

\begin{figure*}[!t]
\centering \subfigure[Loss function vs. the number of
iterations.]{\includegraphics[width=2.3in]{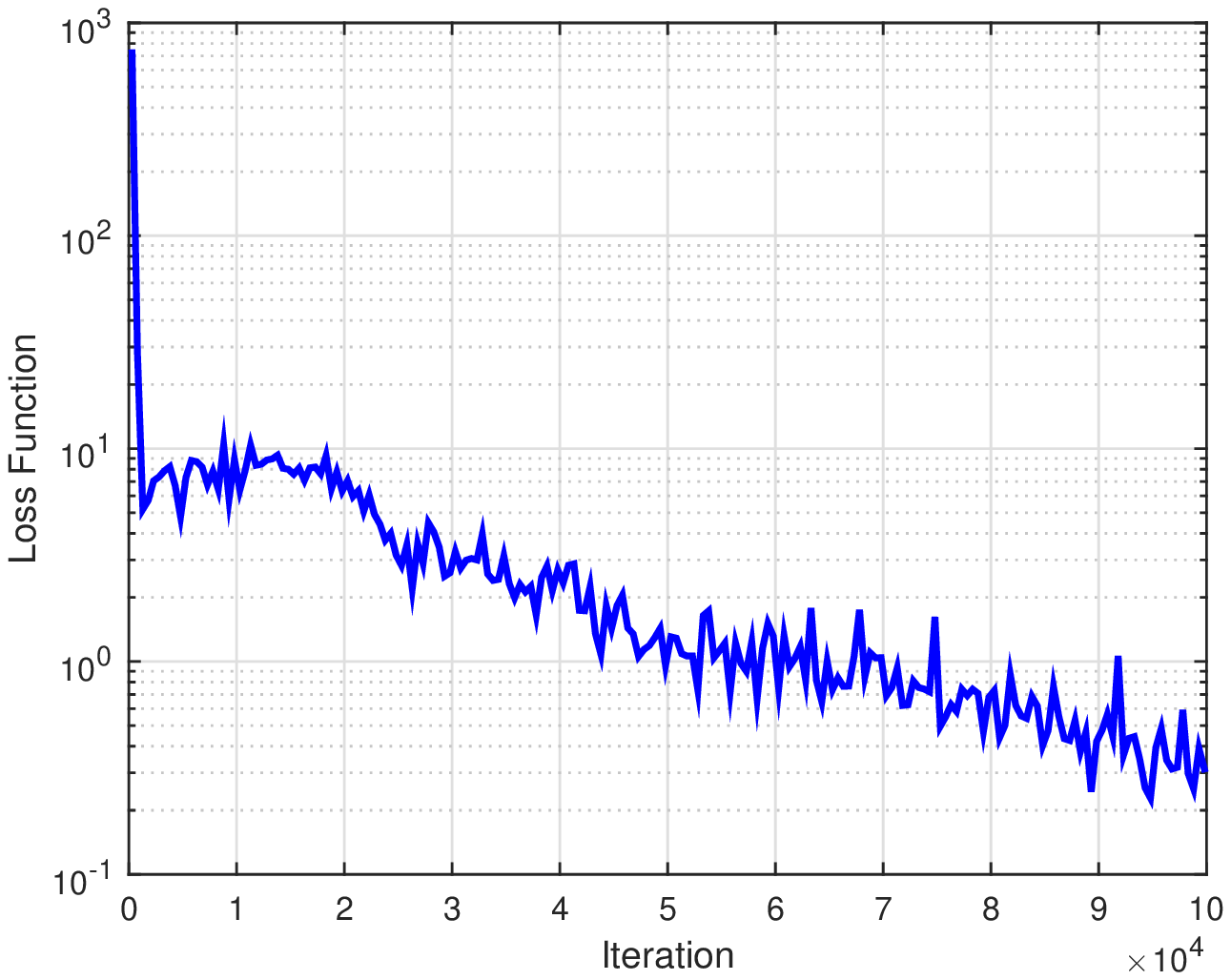}} \hfil
\subfigure[Success rate vs. the number of
iterations.]{\includegraphics[width=2.3in]{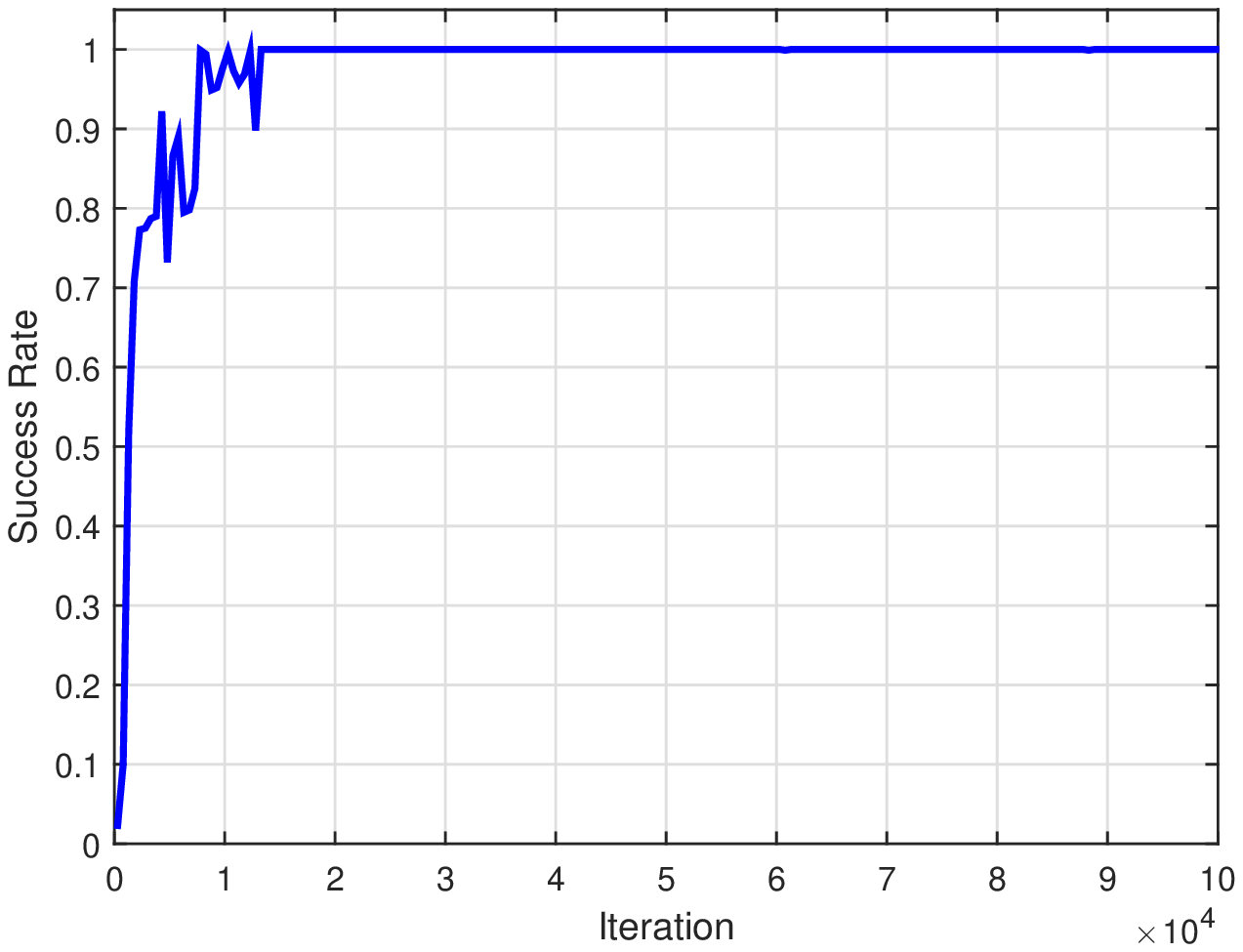}} \hfil
\subfigure[Average number of transition steps vs. the number of
iterations.]{\includegraphics[width=2.3in]{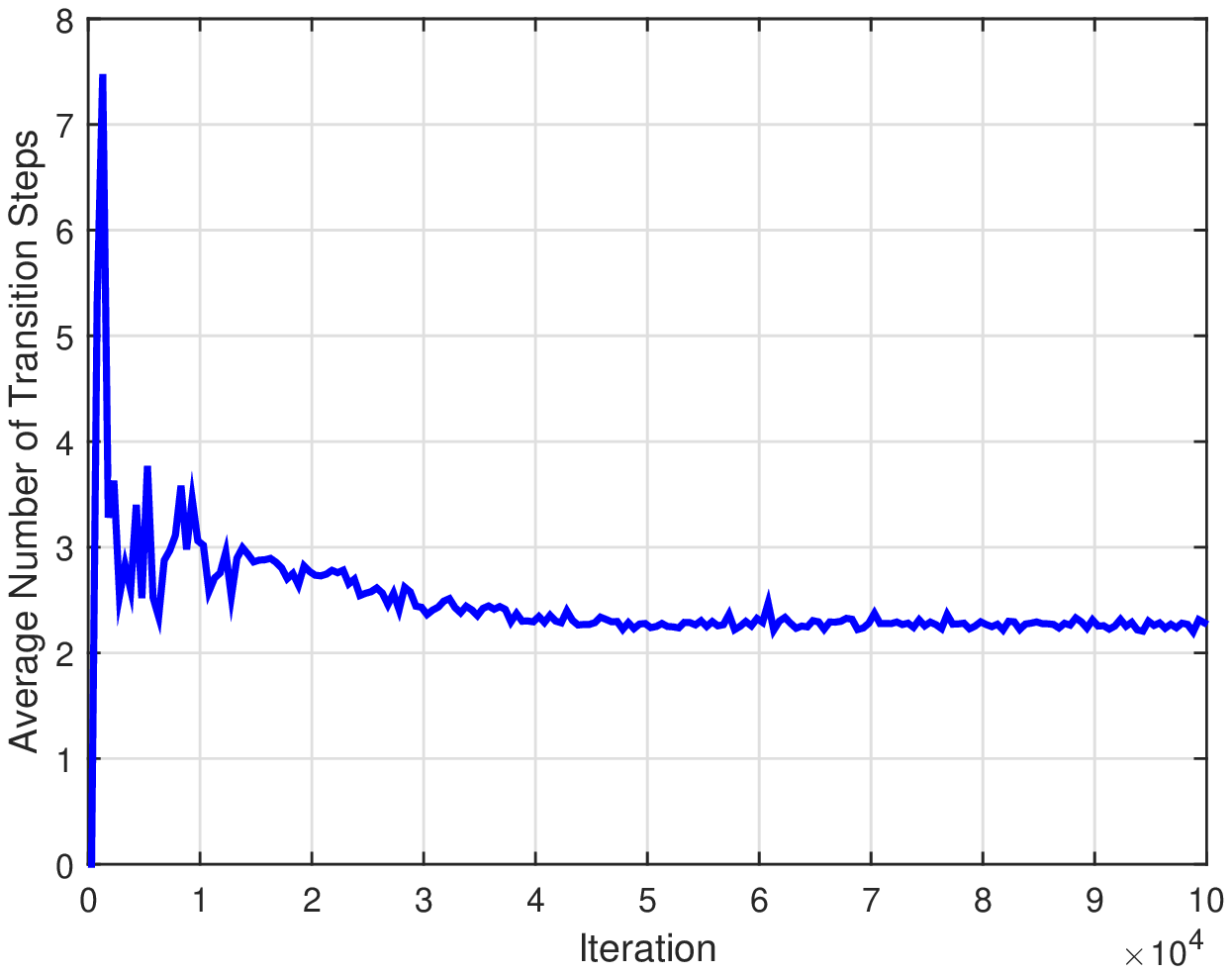}}
\caption{Loss function, success rate, and average number of
transition steps vs. the number of iterations $k$ used for
training, where $N=10$, $\sigma_n = (p_1^pg_{1n}+
p_1^sg_{2n})/10$.} \label{fig21}
\end{figure*}

\begin{figure*}[!t]
\centering \subfigure[Loss function vs. the number of
iterations.]{\includegraphics[width=2.3in]{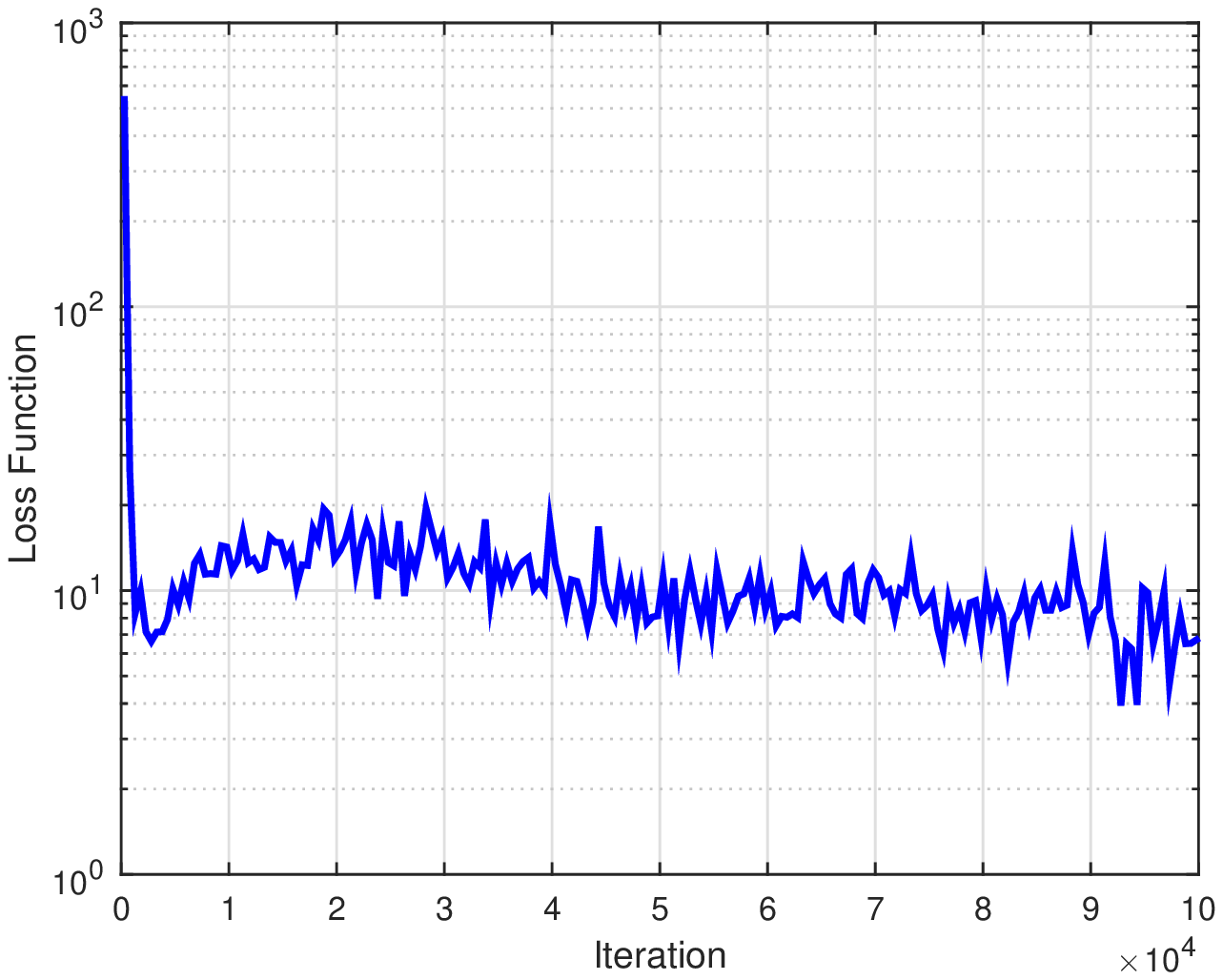}} \hfil
\subfigure[Success rates vs. the number of
iterations.]{\includegraphics[width=2.3in]{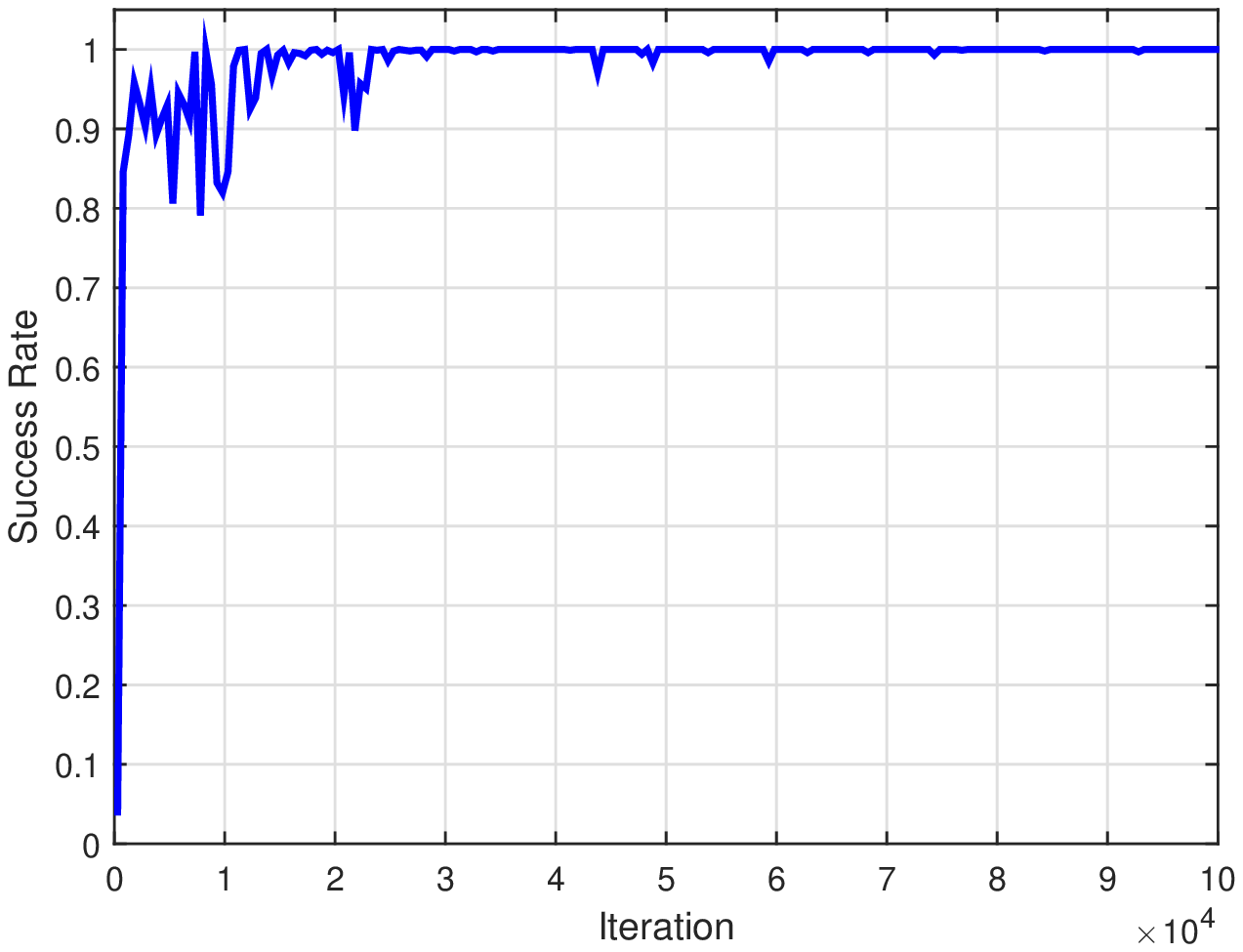}} \hfil
\subfigure[Average number of transition steps vs. the number of
iterations.]{\includegraphics[width=2.3in]{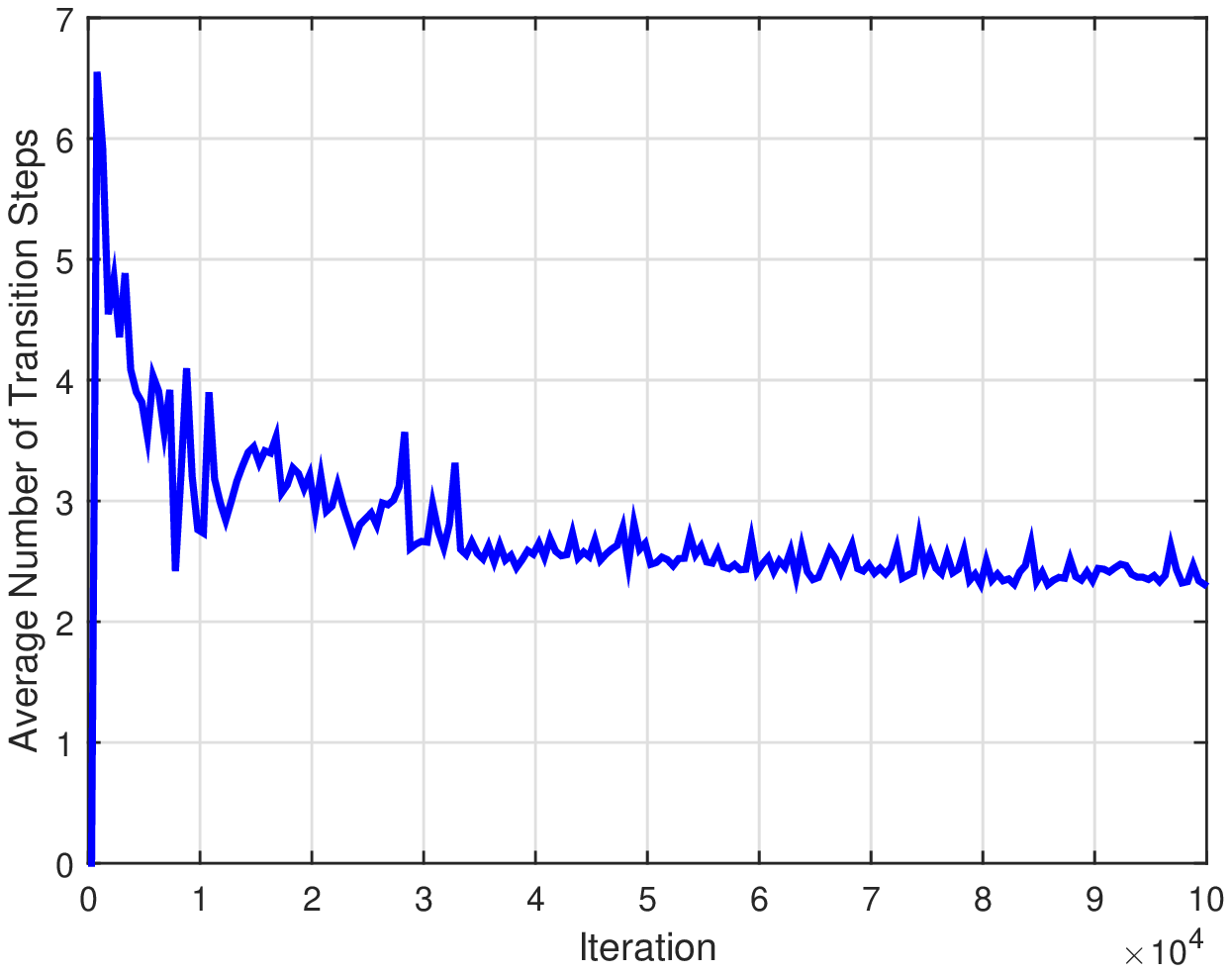}}
 \caption{Loss
function, success rate, and average number of transition steps vs.
the number of iterations $k$ used for training, where $N=10$,
$\sigma_n = (p_1^pg_{1n}+ p_1^sg_{2n})/3$.} \label{fig22}
\end{figure*}

\begin{figure*}[!t]
\centering \subfigure[Loss function vs. the number of
iterations.]{\includegraphics[width=2.3in]{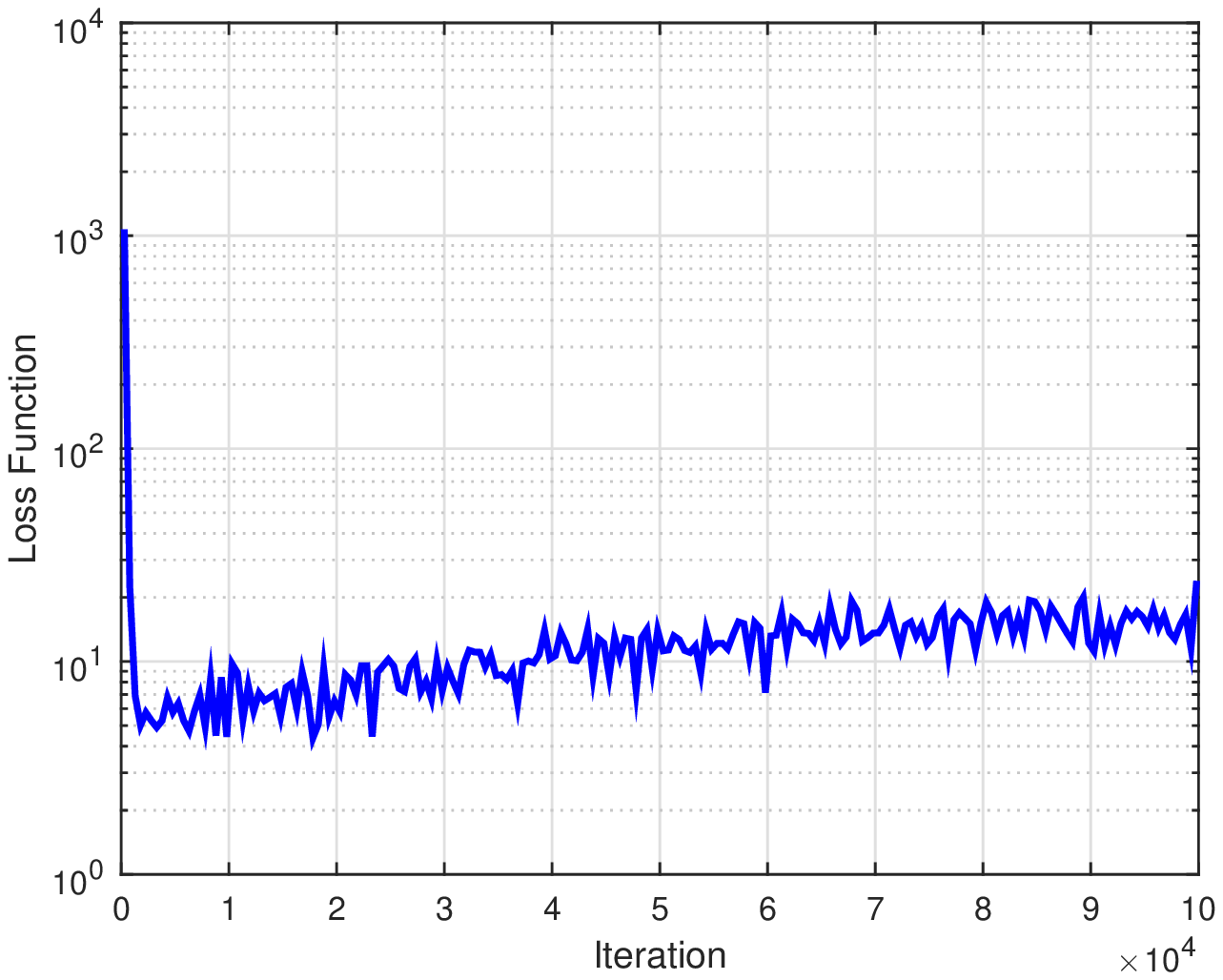}}\hfil
\subfigure[Success rate vs. the number of
iterations.]{\includegraphics[width=2.3in]{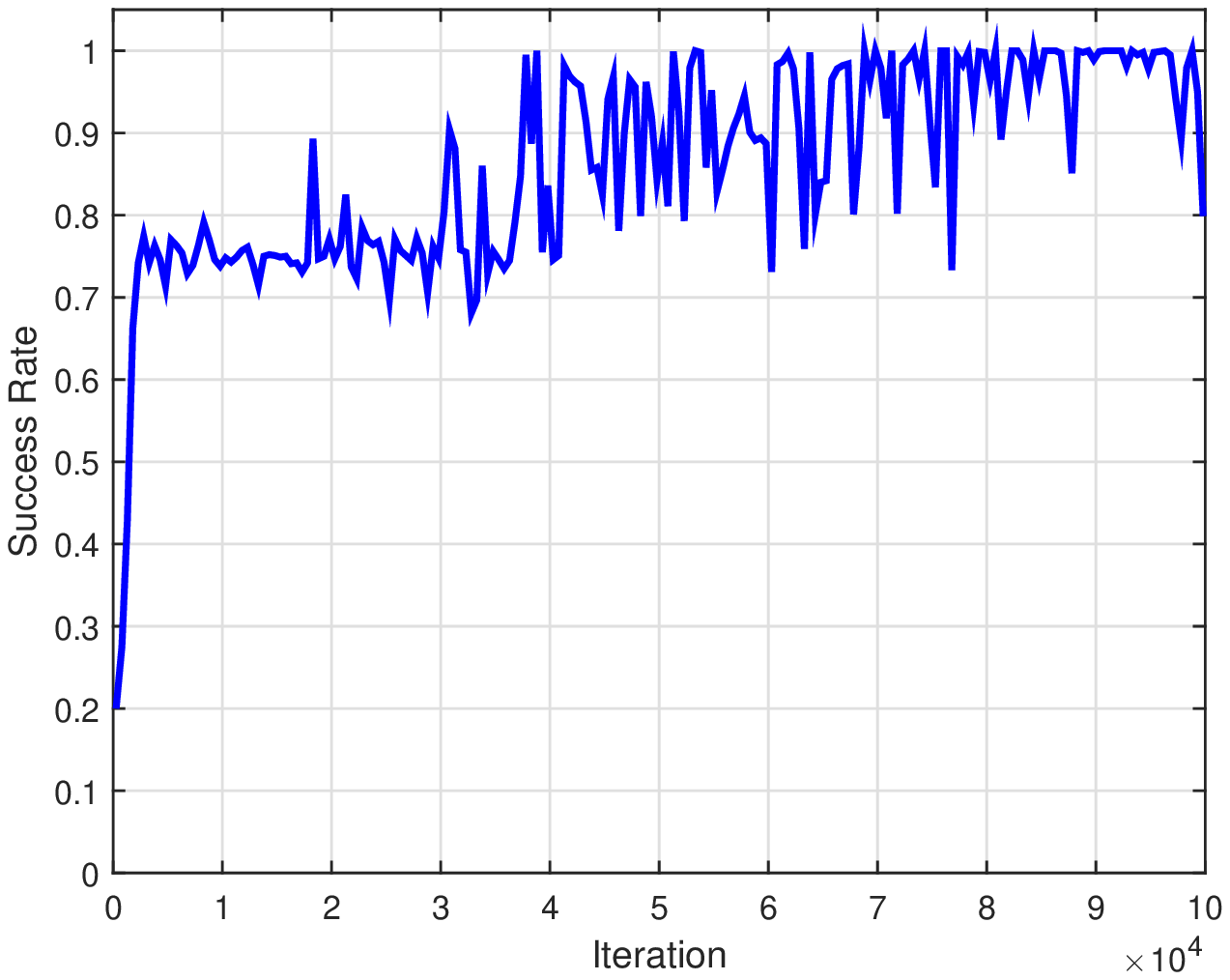}} \hfil
\subfigure[Average number of transition steps vs. the number of
iterations.]{\includegraphics[width=2.3in]{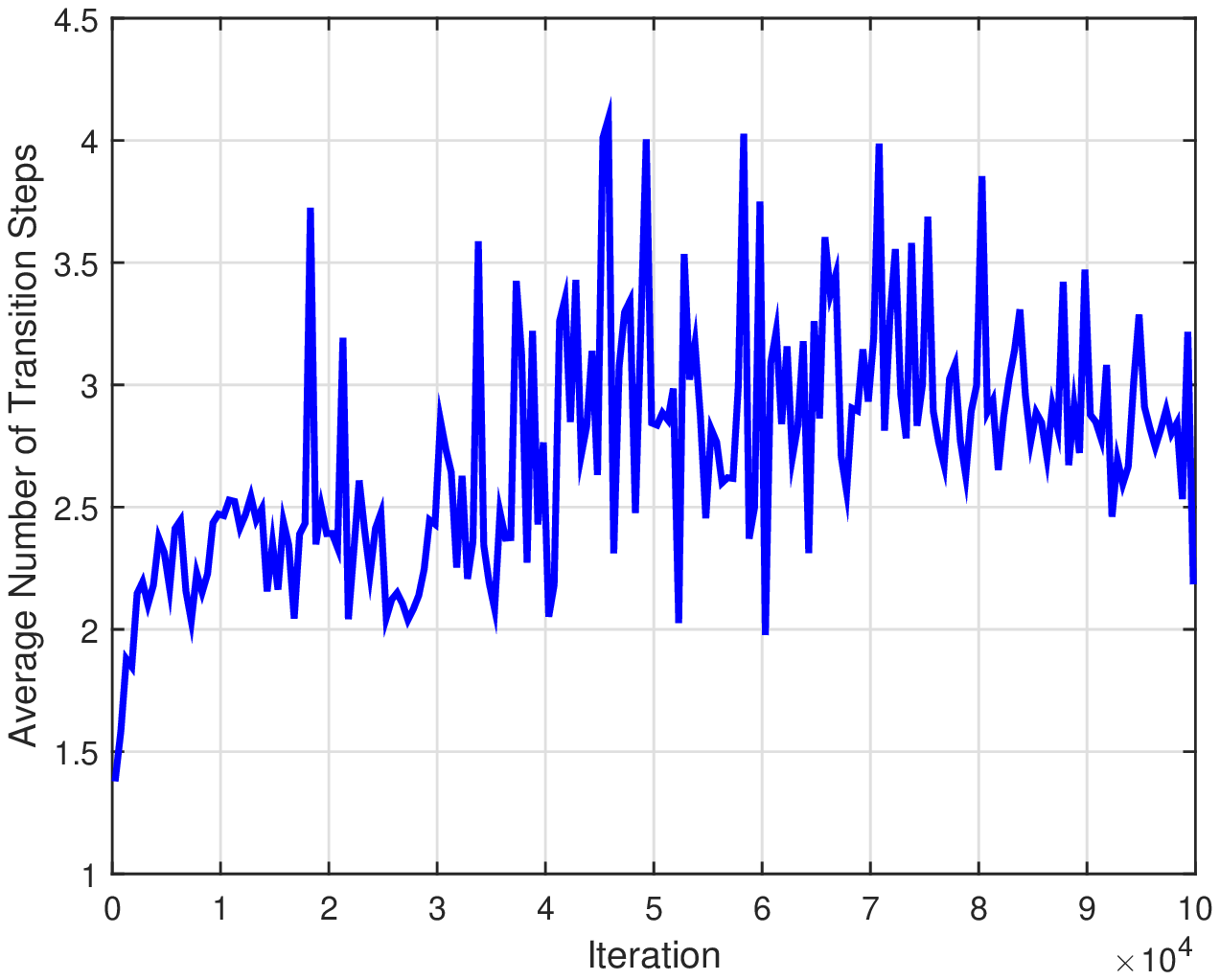}}
\caption{Loss function, success rate, and average number of
transition steps vs. the number of iterations $k$ used for
training, where $N=3$, $\sigma_n = (p_1^pg_{1n}+ p_1^sg_{2n})/10$.
} \label{fig23}
\end{figure*}

In our experiments, the deep neural network (DNN) used to
approximate the action-value function consists of three
fully-connected feedforward hidden layers, and the number of
neurons in the three hidden layers are $256$, $256$, and $512$,
respectively. Rectified linear units (ReLUs) are employed as the
activation function for the first and the second hidden layers. A
ReLU has output 0 if the input is less than 0, and raw output
otherwise. For the last hidden layer, the tanh function is used as
the activation function. The Adam algorithm \cite{KingmaBa14} is
adopted for updating the weights $\boldsymbol{\theta}$, where the
size of a minibatch is set to $256$. We assume that the replay
memory $D$ contains $N_D=400$ most recent transitions, and in each
iteration, the training of $\boldsymbol{\theta}$ begins only when
$D$ stores more than $O=300$ transitions. The total number of
iterations is set to $K=10^5$. The probability of exploring new
actions linearly decreases with the number of iterations from
$0.8$ to $0$. Specifically, at iteration $k$, we let
\begin{align}
\varepsilon_k = 0.8(1-k/K)
\end{align}
We use Algorithm \ref{DQN-training} to train the network, and use
Algorithm \ref{DQN-based-Power-Control} to check its performance.



The performance is evaluated via two metrics, namely, the success
rate and the average number of transition steps. The success rate
is computed as the ratio of the number of successful trials to the
total number of independent runs. A trial is considered successful
if $\boldsymbol{s}$ moves to a goal state within 20 time frames.
The average number of transition steps is defined as the average
number of time frames required to reach a goal state if a trial is
successful.



We now study the performance of the deep reinforcement learning
approach. Specifically, we examine the loss function, the success
rate, and the average number of transition steps as a function of
the number of iterations $k$ used for training. During training,
the loss function is calculated according to
(\ref{loss-function}). After $k$ iterations of training, the
secondary user can use the trained network to interact with the
primary user. The success rate and the average number of
transition steps are used to evaluate how well the network is
trained. Results are averaged over $10^3$ independent runs, in
which a random initial state is selected for each run. Fig.
\ref{fig11} plots the loss function, the success rate, and the
average number of transition steps vs. the number of iterations
$k$, where we set $N=10$, the standard deviation of the random
variable used to account for the shadowing effect and measurement
errors is set to $\sigma_n=(p_1^pg_{1n}+ p_1^sg_{2n})/10$, and the
primary user employs (\ref{power-control-policy-1}) to update its
transmit power. We see that the secondary user, after only $10^3$
iterations of training, can learn an efficient power control
policy which ensures that a goal state can be reached quickly
(with $1.5$ average number of transition steps) from any initial
states with probability one. Fig. \ref{fig12} and Fig. \ref{fig13}
depict the loss function, the success rate, and the average number
of transition steps vs. $k$ for different choices of $N$ and
$\sigma_n$, where we set $N=10$, $\sigma_n = (p_1^pg_{1n}+
p_1^sg_{2n})/3$ for Fig. \ref{fig12} and $N=3$,
$\sigma_n=(p_1^pg_{1n}+ p_1^sg_{2n})/10$ for Fig. \ref{fig13}. We
see that the value of the loss function becomes larger when we
increase the variance $\sigma_n$ or decrease the number of
sensors. Nevertheless, the learned policy is still very efficient
and effective, attaining a success rate and an average number of
transition steps similar to those in Fig. \ref{fig11}. This result
demonstrates the robustness of the deep reinforcement learning
approach.

Next, we examine the performance of the DQN-based power control
method when the primary user employs the second power control
policy (\ref{power-control-policy-2}) to update its transmit
power. Since the policy (\ref{power-control-policy-2}) is more
conservative, the task of learning an optimal power control
strategy is more challenging. Fig. \ref{fig21} depicts the loss
function, the success rate, and the average number of transition
step as a function of $k$, where we set $N=10$ and
$\sigma_n=(p_1^pg_{1n}+ p_1^sg_{2n})/10$. We observe that for this
example, more iterations (about $1.5\times10^4$) are required for
training to reach a success rate of one. Moreover, the learned
policy requires an average number of transition steps of $2.5$ to
reach a goal state. The increased number of transition steps is
because the second policy used by the primary user only allow its
transmit power to increase/decrease by a single level at each
step. Thus more steps are needed to reach the goal state. Fig.
\ref{fig22} and Fig. \ref{fig23} plot the loss function, the
success rate, and the average number of transition steps vs. $k$
for different choices of $N$ and $\sigma_n$, where we set $N=10$,
$\sigma_n = (p_1^pg_{1n}+ p_1^sg_{2n})/3$ for Fig. \ref{fig22} and
$N=3$, $\sigma_n = (p_1^pg_{1n}+ p_1^sg_{2n})/10$ for Fig.
\ref{fig23}. For this example, we see that a large variance in the
state observations and an insufficient number of sensors lead to
performance degradation. In particular, the proposed method incurs
a considerable performance loss when fewer sensors are deployed.
This is because the random variation in the state observations
makes different states less distinguishable from each other and
prevents the agent from learning an effective policy, but using
more sensors helps neutralize the effect of random variations.



\begin{figure}[!t]
\centering
\includegraphics[width=3.5in]{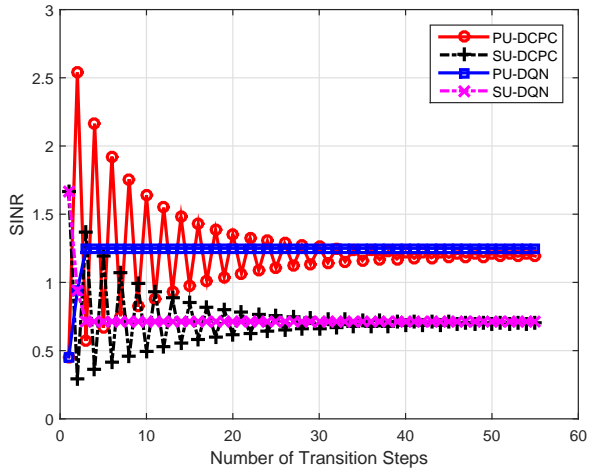}
\caption{SINRs of the primary and secondary users vs. the number
of transition steps.} \label{SINRvsiter}
\end{figure}

Lastly, we compare the DQN-based power control method with the
DCPC algorithm \cite{GrandhiZander94} which was developed for
power control in an optimization framework. For the DCPC
algorithm, the primary user and secondary user use the following
power control policy to update their respective transmit power:
\begin{align}
  p_1(k+1) = \min\left\{p^p_{L_1}, \frac{\eta_1p_1(k)}{\text{SINR}_1(k)}\right\}
\label{DCPC-power-control-policy-1}
\end{align}
\begin{align}
  p_2(k+1) = \min\left\{p^s_{L_2}, \frac{\eta_2p_2(k)}{\text{SINR}_2(k)}\right\}
\label{DCPC-power-control-policy-2}
\end{align}
For the DQN-based method, the primary user uses the policy
(\ref{power-control-policy-1}) to update its transmit power, the
number of sensor nodes and the state observation noise variance
are set to $N=10$ and $\sigma_n=(p_1^pg_{1n}+ p_1^sg_{2n})/10$,
respectively. In Fig. \ref{SINRvsiter}, we examine the QoSs (i.e.
SINRs) of the primary and secondary users as the iterative process
evolves. We see that although both schemes can converge from an
initial point, our proposed DQN-based method requires only a few
transition steps to reach a goal state, while the DCPC algorithm
takes tens of steps to converge. We also observe that the
DQN-based scheme converges to a solution that is close to the
optimal solution obtained by the DCPC algorithm, which further
corroborates the effectiveness of the proposed DQN-based scheme.
Note that optimization-based techniques such as the DCPC algorithm
require global coordination among all users in the cognitive
networks so that the primary user and the secondary user can
interact in a cooperative way. In contrast, for our proposed
scheme, the primary user follows its own rule to react to the
environment. In other words, the interaction between the primary
user and the secondary user is not planned out in advance and
needs to be learned in real time. Although the training of the DQN
involves a high computational complexity, after the training is
completed, the operation of the power control has a very low
computational complexity: given an input state $\boldsymbol{s}$,
the secondary user can make a decision using simple calculations.


\section{Conclusions} \label{sec:conclusions}
We studied the problem of spectrum sharing in a cognitive radio
system consisting of a primary user and a secondary user. We
assume that the primary user and the secondary user work in a
non-cooperative way. The primary user adjusts its transmit power
based on its own pre-defined power control policy. We developed a
deep reinforcement learning-based method for the secondary user to
learn how to adjust its transmit power such that eventually both
the primary user and the secondary user are able to transmit their
respective data successfully with required qualities of service.
Experimental results show that the proposed learning method is
robust against the random variation in the state observations, and
a goal state can be reached from any initial states within only a
few number of steps.

\bibliography{newbib}
\bibliographystyle{IEEEtran}

\end{document}